\documentstyle[twocolumn,psfig,rotating]{mn2e}
\def\lsim{~\rlap{$<$}{\lower 1.0ex\hbox{$\sim$}}}
\def\bsim{~\rlap{$>$}{\lower 1.0ex\hbox{$\sim$}}}
\def\yr{\ {\rm yr}}
\def\erg{\,{\rm erg}}
\def\ergscms{\,{\rm erg\,cm^{3}\,s^{-1}}}
\def\kms{\ {\rm km\,s^{-1}}}
\def\hmpc{\ {\rm h^{-1}Mpc}}
\def\hkpc{\ {\rm h^{-1}kpc}}
\def\mdh{\ {\rm M_\odot/h}}
\def\hhmpc{\ {\rm h^{2}Mpc^{-3}}}
\def\hhhmpc{\ {\rm h^{3}Mpc^{-3}}}
\def\myr{\ {\rm M_\odot yr^{-1}}}
\def\cmm{\ {\rm cm^{2}}}
\def\pcc{\,{\rm cm^{-3}}}
\def\kel{\,{\rm K}}

\def\la{\langle}
\def\ra{\rangle}
\def\rw{{r_{\rm w}}}
\def\vw{{v_{\rm w}}}
\def\rigm{{\rho_{\rm IGM}}}
\def\vx{{\bf x}}
\def\rhof{{\rho^{\rm s}_{_{\rm dm}}}}
\def\vf{{{\bf v}^{\rm s}_{_{\rm dm}}}}
\def\dhi{n_{_{\rm HI}}}
\def\vhi{v_{_{\rm p}}}
\def\dhihat{{\hat n}_{_{\rm HI}}}
\def\xihf{\xi_{\rm hf}}
\def\om#1{\Omega_{#1}}
\def\etal{{\it et al.\ }}
\newcommand{\op}{Ly$\alpha$\ }
\newcommand{\hi}{\mbox{H{\scriptsize I}}}

\def\lya{\mbox{Ly$\alpha$\phantom{ }}}

\begin{document}
\title[]
{Galactic winds and the \lya forest}
\author[Desjacques {\it et al.}]{Vincent Desjacques$^1$, Adi Nusser$^1$,
Martin G. Haehnelt$^2$,
\newauthor{ and Felix Stoehr$^3$}\\
$^1$Physics Department and Space Research Institute, Technion-Haifa\\
$^2$Institute of Astronomy, Madingley Rd, CB3-0HA, Cambridge, UK\\
$^3$Max-Planck-Institut f\"ur Astrophysik, Karl-Schwarzschild-Str. 1, 
D--85740 Garching b. M\"unchen, Germany}
\maketitle

\begin{abstract}

We study the effect of galactic outflows on the statistical properties
of  the \op\ forest and its correlation with galaxies. The winds are
modelled as fully ionised spherical bubbles centered around the haloes
in an  N-body simulation of a $\Lambda$CDM model. The observed flux 
probability distribution and flux  power spectrum limit the 
volume filling factor of  bubbles to be less than  $10\%$.   
We have compared the mean flux as a function of distance from haloes
with the Adelberger \etal (ASSP) measurement. For a model of bubbles
of constant size surrounding the most massive haloes, bubble radii of
$\bsim 1.5\hmpc$ are necessary to match the high  transmissivity at
separations $\leq 0.5\hmpc$ but the increase of the transmissivity at
small scales is more gradual than observed. The cosmic variance error
due  to the finite number of galaxies in the sample increases rapidly
with  decreasing separation.  At separations $\leq 0.5 \hmpc$  our
estimate  of the cosmic variance error is  $\Delta{\bar F}\sim 0.3$,
30\% higher than  that of ASSP.  The difficulty in matching the rise
in the transmissivity at  separations smaller than the size of the
fully ionised bubbles surrounding the  haloes is caused by residual
absorption of neutral hydrogen lying physically  outside the bubbles
but having a redshift position similar to the haloes. The  flux level
is thus sensitive to the amplitude  of the
coherent velocity shear near halos and to a smaller extent to 
the amplitude of thermal motions. We find that the velocity shear
increases with halo mass in the simulation. A model where  LBGs are
starbursts in small mass haloes matches the observations with smaller
bubble radii than a model where massive haloes host the LBGs.  If we
account for the uncertainty in the redshift position of haloes, a
starburst model  with a bubble  radius of $1\hmpc$ and a volume
filling  factor of 2\% is consistent with the ASSP measurements at the
1-1.5$\sigma$ level.  If this model is correct the sharp rise of the
transmissivity at separations  $\leq 0.5 \hmpc$ in the ASSP sample is
due to cosmic variance and is expected to become more moderate for a
larger sample.

\end{abstract}


\section {Introduction}
\label{introduction}

Stellar feedback, in the form of ionising radiation and mechanical
energy in galactic winds/outflows, plays an important role in
determining the physical properties of galaxies and the Intergalactic
Medium (IGM). The galactic contribution to the hydrogen ionising 
UV radiation may well exceed that of QSOs. Galactic superwinds
regulate star formation, enrich the IGM with metals, and alter the
dynamical and observational properties of the IGM close to galaxies.

Galactic winds are driven by the mechanical energy produced by
supernova explosions of massive stars. They have  been studied
extensively in nearby starburst galaxies ({\it e.g. } Heckman \etal
1990). The large velocity shifts between stellar and interstellar
lines in the spectra of Lyman-break galaxies (LBGs) are strong
evidence that  galactic winds  are also present in  high-redshift
galaxies  (Pettini \etal 1998, 2000, Heckman \etal 2000).  The
inferred  wind velocities are several hundred  $\kms$, far  above the
sound speed of  the IGM so that strong shocks are likely to be
associated with these winds. These shocks will  heat up, collisionally
ionise, and sweep up  the surrounding IGM near galaxies (Heckman
2000).   Adelberger \etal (2003, hereafter ASSP) have analyzed a
sample of high  resolution QSO spectra with lines-of-sight (LOS)
passing very close   to LBGs. They detected a significant decrease of
the \op\ absorption near galaxies. As discussed by ASSP and Kollmeier
et al. (2003) the increased UV flux close to LBGs falls short 
of explaining the decreased absorption by a factor of a few.  
The decreased absorption is thus taken as  evidence for the existence 
of dilute and highly ionised gas bubbles caused by wind-driven shocks. 

Initially a shock propagating in the IGM expands adiabatically.
During this phase the gas encountered by the shock is heated and
collisionally ionised to become part of a bubble of dilute and hot
plasma.  When radiative losses become important, a thin cool shell
forms at the shock front.  The shell is driven outwards by the hot low
density inner plasma (e.g. Weaver \etal 1977, Ostriker \& McKee 1988,
Tegmark \etal 1993).  Eventually the expansion of the plasma comes to
a halt when pressure equilibrium with the surrounding IGM is reached.
To explain the lack of \hi\ close to the haloes, the bubble has to
survive for a sufficiently long time.  We defer the discussion of the
conditions for the survival of the bubble to another paper.  Here we
assume that the bubbles are long-lived and study the  implications of
long-lived bubbles devoid of neutral hydrogen on QSO absorption
spectra.

We will investigate how the correlation between the flux in the  \op\
forest and the distance of the LOS to the closest galaxy depends   on
physical  properties of the ionised bubbles around  galaxies. For this
purpose we use a N-body simulation of  collisionless dark matter
(DM). The basic observed properties of the \op\ forest can be
reproduced with such simulations using a simple  prescription that
relates the distribution of gas and dark matter in  the IGM  (Hui \&
Gnedin 1997). With DM simulations it is possible  to probe a wider
dynamical range than with  hydrodynamical simulations. This allows a
better  modelling of the statistical properties of the IGM-halo
relation.

Measures of the halo-flux correlation,  defined for example as the
mean flux in the \op\ forest as  a function of the distance to the
nearest galaxies, are  sensitive to the  size of the bubbles.
Absorption spectra  probe the gas distribution  in redshift space.
Thermal motions and coherent peculiar velocities therefore strongly
affect  the halo-flux correlation at small separations
(e.g. Kollmeier \etal 2003).  Previous modelling of the impact of
winds on  the \op forest  had difficulties to match  the large
decrease of the absorption  measured by ASSP  at comoving separation
$s\leq 0.5\hmpc$  from the galaxies (Croft \etal 2002, Bruscoli \etal
2002,  Theuns \etal 2002, Kollmeier \etal 2003).  We will investigate
here the role of peculiar velocities, thermal motions and the scaling
of bubble size with halo mass in more detail.  We will also perform a
detailed assessment of the errors to obtain a more robust lower limit
on the  size  and thus the volume filling factor of the bubbles.

The paper is organized as follows. In \S\ref{lyman} we describe the
model of the Ly$\alpha$ forest which we have used to calculate
synthetic spectra from the DM simulations.  In \S\ref{nbody}, we
describe the  simulation and the corresponding halo catalogs, compute
three-dimensional flux maps  of the simulation, and  discuss the flux
probability distribution function and power spectrum in the presence
of winds.  The main body of the paper are \S\ref{xihf} and
\S\ref{fluxsim} where we study in detail the impact of galactic winds
on the halo-flux correlation and the corresponding observational
errors. We conclude with a discussion of the results in
\S\ref{discussion}.

\section{The Lyman $\alpha$ Forest and the Moderate Density IGM}
\label{lyman}

The \op forest is now widely believed to originate from  an undulating
warm ($~10^{4}$K) photoionized IGM.  The \op optical depth in redshift
space due to resonant scattering can be expressed as a convolution of
the real space \hi\ density along the line of sight with a Voigt
profile ${\cal H}$ (Gunn \& Peterson 1965, Bahcall \& Salpeter 1965),
\begin{eqnarray}
\tau(w) &=& \frac{c \sigma_0}{H(z)}\nonumber \\
& & \times \int_{-\infty}^{+\infty}\dhi(x){\cal H} \left[w-x-\vhi(x),b(x) 
\right] dx\; .
\label{depth}
\end{eqnarray}
where $\sigma_0=4.45\times 10^{-18}\cmm$ is the effective
cross-section for resonant line scattering and H$(z)$ is the Hubble
constant at redshift $z$, $x$ is the real space coordinate, $v_p(x)$
is the  line of sight component of the \hi\ peculiar velocity field,
${\cal H}$ is the Voigt profile, and $b(x)$ is the Doppler parameter
due to  thermal/turbulent broadening. For thermal broadening
$b(\vx)=13\left({{T(\vx)}\over 10^4\kel}\right)^{1/2}\kms$, where T is
the temperature of the gas.

Both $x$ and $w$ are measured in $\kms$.  The Voigt profile is
normalized such that $\int{\cal H}=1$ and can be  approximated by a
Gaussian for moderate optical depths, ${\cal H}=1/\left(\pi^{1/2}
b\right)\times\exp\left\{-[\omega-x-v_p(x)]^2/b^2\right\}$.

The normalized flux obtained by ``continuum fitting'' of the observed
spectrum is  related to the optical depth along the line of sight as, 
\begin{equation}                                                            
F(w)\equiv I_{\rm obs}(w)/I_{\rm cont}={\rm e}^{-\tau(w)}\; ,
\end{equation}
where $\tau$ is the optical depth, $w$ is the redshift space coordinate 
along the line of sight, $I_{\rm obs}$ is the observed flux and 
$I_{\rm cont}$ is the flux emitted from the source (quasar) that would 
be observed in the absence of any intervening material.

On scales larger than a filtering scale which is related to the  Jeans
scale , $x_{\rm J}$, the IGM traces the DM distribution very well.
For moderate overdensities the balance between  adiabatic cooling and
photo-heating of the expanding IGM  establishes  power-law relations
between  temperature, neutral hydrogen density and total gas density
(Katz \etal 1996, Hui \& Gnedin 1997, Theuns \etal 1998),
\begin{equation} T_{\rm g}\propto \rho_{\rm g}^\beta \quad {\rm and}\quad 
\dhi\propto \rho_{\rm g}^\alpha \; ,
\label{state}
\end{equation} 
where the parameter $\beta$ is in the range $0-0.62$,
$\alpha=2-0.7\beta$, and $\rho_{\rm g}$ and $T_{\rm g}$ are the
density and temperature of the gas, respectively.

On scales smaller than the Jeans scale pressure dominates over
gravity.  The gas pressure smoothes  the gas distribution compared to
the distribution of collisionless dark matter ({\it e.g.}  Theuns,
Schaye \& Haehnelt 2000). We thus  assume that $\rho_{\rm
g}(\vx)\propto \rhof(\vx)$,  where $\rhof(\vx)$ is the dark matter
density smoothed in a way  that mimics the pressure effects.  We adopt
the smoothing used  in SPH simulations (e.g. Springel, Yoshida \&
White 2001)  to estimate the gas density from the dark matter particle
distribution.  The \hi\ density $\dhi$ and the smoothed dark matter
field $\rhof$ are then related as
\begin{equation}
\dhi(\vx)= \dhihat \left[\frac{\rhof(\vx)}{{\bar \rho}_{\rm dm}^{\rm
 s}}\right]^{\alpha}\; ,
\label{neutral}
\end{equation}
where ${\bar \rho}_{\rm dm}^{\rm s}$ is the (volume) average of
$\rhof$ , and $\dhihat$ is the \hi\ density at $\rhof={\bar
\rho}_{\rm dm}^{\rm s} $.

\section{Simulating galaxies and the flux distribution}
\label{nbody}

\subsection{The DM simulation}  

One of our main goals is to assess the error in the observed  galaxy
flux correlation. This requires a simulation  large  enough to contain
a substantial number of DM haloes which can  be identified as hosts of
LBGs.  At the same time the simulation should still reproduce typical
absorption systems.  This favours the use of a DM  simulation for
which a larger  dynamical range can be achieved rather than  a
computationally  more expensive  hydrodynamical simulation.  We use a
$\Lambda$CDM  simulation  run  with  the N-body code GADGET (Springel,
Yoshida  \& White 2001). The cosmological parameters are listed in
Table~\ref{table1}.  It is part of a series of simulations of
increasingly higher resolution which zoom in  on a  spherical region
of initial comoving radius 26$\hmpc$  selected from a simulation 
with box size 479$\hmpc$ (cf. Stoehr \etal 2002).  The
inner region of the simulation  has thus higher resolution than the
outer regions.  We only use the  particle distribution inside a box
of size $L=30\hmpc$  which belongs to the spherical region of high
resolution.  The particle mass in this  high-resolution region is
$m=9.52\times 10^8\mdh$ (cf. Table~\ref{table1}).  The large volume
makes  the simulation well suited for  a direct  comparison   with
the observed galaxy-flux correlation.  We use a redshift output $z=3$
to facilitate a comparison with the ASSP  measurements.

\begin{table}
\caption{The principle parameters of the N-body simulation.   
The number of particles, $N$,  and the particle mass,
$m$, in the high resolution region are also listed. }
\vspace{1mm}
\begin{center}
\begin{tabular}{cccccc} 
\hline
$\om{\rm m}$ & $\om{\Lambda}$ & h & $\sigma_8$&N&$m/[10^8\mdh]$ \\
\hline
0.3  & 0.7 & 0.7 & 0.9& 12067979 &  9.52 \\
\hline\hline
\end{tabular}
\end{center}
\label{table1}
\end{table}

\subsection{DM haloes as host of LBGs}

We aim at extracting halo catalogs with statistical properties similar
to those of observed  Lyman-break galaxies (LBGs). However, the
relation between LBG and DM haloes is still  somewhat uncertain.  When
LBGs were first discovered, a simple picture was advocated in which
the mass of the DM haloes hosting the LBGs scales approximately
linearly with the luminosity of LBGs, LBGs are long-lived and each DM
halo hosts  one LBG (e.g. Steidel \etal 1996; Adelberger \etal 1998,
Bagla 1998; Haehnelt, Natarajan \& Rees 1998).  In this
``massive-halo'' scenario the masses of the host haloes are rather
large, $M\bsim 10^{12}\mdh$, and the high-redshift LBGs are the
progenitors of today's massive and luminous galaxies.  The rather slow
decrease of the  LBG space density with  increasing redshift motivated
an alternative picture,  where most of the observed LBGs are
interacting starburst galaxies  in the process of assembling
(e.g. Lowenthal \etal 1997, Trager \etal 1997, Kolatt \etal 1999,
Somerville \etal 2001).  We will call this scenario the ``starburst''
picture in the following.  It predicts that most  of the observed LBGs
are hosted by haloes  of smaller  mass $M \sim 10^{11}\mdh$, which
will eventually merge to form more typical galaxies at $z\simeq
0$. Both scenarios predict similar  clustering properties and cannot
be ruled out based on the available  clustering data (e.g. Wechsler
\etal 2001).  Direct observational  constraints on the masses of LBGs
are also still rather weak.  Studies of rest-frame optical nebular
emission lines imply halo masses in the range $10^{10}-10^{11}\mdh$
(e.g. Pettini \etal 1998), while near-infrared imaging surveys
combined with stellar population synthesis models and extinction yield
virial masses of $M\sim 10^{11}-10^{12}\mdh$  (e.g. Shapley \etal
2001).     We will here first adopt  the massive halo picture {\it
i.e.}    we assume that a dark matter halo contains  only one LBG
(e.g. Adelberger \etal 1998) and pick a minimum mass $M_{\rm min}$
such that the number density of haloes $M\geq M_{\rm min}$ in the
simulation is similar to the observed number density of high-redshift
LBGs.   In Section~\S\ref{m:role} we will  discuss what happens if we
relax these assumptions.

\begin{figure}
\mbox{\psfig{figure=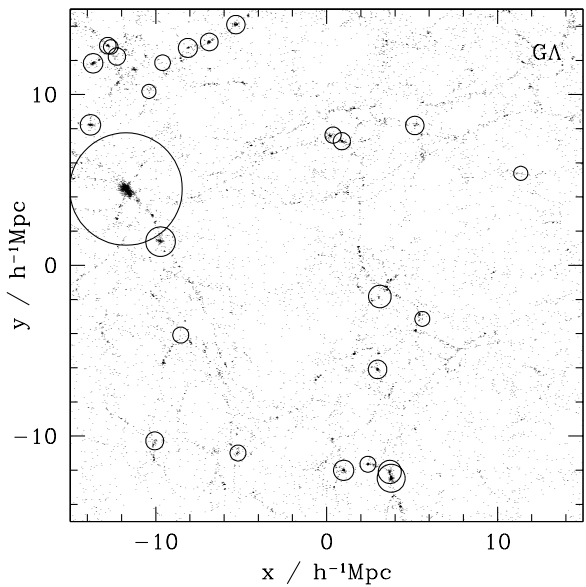,height=3.2in,width=3.2in}}
\caption{The particle distribution at  $z=3$ in a slice of thickness
$\Delta h=0.21\hmpc$.  Haloes with masses $M\geq 5\times 10^{10}\mdh$
are  represented  by the open circles with radii proportional to their
masses.}
\label{fig1}
\end{figure}

DM haloes were identified with a Friends-of-Friends group finding
algorithm. Only groups containing at least 20 dark matter (DM)
particles were classified as haloes. Since the DM particle mass is
$9.52\times 10^8\mdh$, the minimum halo mass is about $2\times
10^{10}\mdh$.  Figure~\ref{fig1} shows a slice of thickness $\Delta
h=0.21\hmpc$ at  $z=3$ in the simulation. The centers of the open
circles show the  location of haloes with $M\geq 5\times 10^{10}\mdh$.
The radii  of the circles are proportional to their masses.

In the simulation the number density of haloes with $M\geq 10^{11}\mdh$
is about $0.04\hhmpc$, whereas for haloes with $M\geq 10^{12}\mdh$ it
is about $2\times 10^{-3}\hhmpc$ (cf. Table~\ref{table2}).   We have
chosen $M_{\rm min}=5\times 10^{11}\mdh$ as our reference  value but
have also studied alternative choices.  The corresponding number
density is $4\times 10^{-3}\hhmpc$, which is about the number density
of LBGs as determined for a $\Lambda$CDM cosmology at $z=3$ from
high-redshift surveys with limiting magnitude $25\leq {\cal R}\leq 27$
(Giavalisco \& Dickinson 2001). It should be noted that the halo
cumulative function  $n_{\rm h}(m_{\rm h})$ as shown in
Table~\ref{table2} is consistent with a Sheth-Tormen mass function
computed for a $\Lambda$CDM Universe with identical parameters as our
simulation (cf. Table~\ref{table1}).

\begin{table}
\caption{Number densities of haloes  with mass greater than the listed
four values. }
\vspace{1mm}
\begin{center}
\begin{tabular}{ccccc} \hline
       &  &    &  & \\
       $m_{\rm h}/[10^{10}\mdh]$& 5 &10 &50 &$10^2$  \\
\hline
     $n_{\rm h}(>m_{\rm h})/[10^{-3}\hhhmpc]$& 81 & 36 & 3.9 & 1.6 \\
\hline\hline
\end{tabular}
\end{center}
\label{table2}
\end{table}

\subsection{Calculating synthetic spectra from the dark matter 
distribution}
  
\subsubsection{From density   to flux}

The \hi\ and the flux distributions are calculated  from the DM
distribution which was smoothed  with  a SPH kernel containing $20$
particles within  the smoothing length.  The smoothed   DM density
and velocity fields, $\rhof$ and $\vf$,   were interpolated  for
randomly chosen LOS through the simulation box. The pixel size of the
spectra  is $\Delta=60\hkpc$ and the comoving length $30\hmpc$.   From
the density and velocity fields along the LOS  we computed the optical
depth  as a function  of the redshift space  coordinate $w=x+\vhi$,
where $\vhi$ is the component of $\vf$ in the line  of sight
direction.   Eq.~(\ref{depth}) then takes the  following  discrete
form,
\begin{equation}
\tau(w_i)={\cal A}(z)\Delta\sum_j\left[\rhof(x_j)\right]^{\alpha} 
{\cal H}_{ij}\; ,
\label{depthsim}
\end{equation}
where  ${\cal A}=c\sigma_0  n_{\rm HI} ({{\bar \rho}_{\rm dm}^{\rm s}})/H(z)$ 
and we assume that $\rhof$ is normalized such that its mean is unity
and ${\cal H}_{ij}={\cal H}[w_i-x_j-\vhi(x_j),b(x_j)]$ is a normalized
Gaussian profile. 

\subsubsection{The temperature distribution} 

As we are using a DM simulation we still have to specify a
temperature.  For this we use the fact that temperature and density
are tightly  coupled. At moderate densities a simple power-law
relation holds,   $T_{\rm g}\sim\rhof^\beta$ (cf. eq~\ref{state}). In
high-density regions, line cooling is efficient and the gas rapidly
cools  to  $T\sim 10^4\kel$ ({\it e.g.} Theuns \etal 1998,  Dav\'e
\etal 1999).  We will thus take the following  relation between
temperature and density
\begin{equation}
T_{\rm g}=\left\lbrace
\begin{array}{cc}
{\hat T}\left(\frac{\rhof}{{\bar \rho}_{\rm dm}^{\rm
 s}}\right)^\beta & 
\frac{\delta\rho}{\rho}\leq 50 \\
10^4\kel & \frac{\delta\rho}{\rho}>50
 \end{array}\right. \;
\label{statesim}
\end{equation}
In the high-density regions where shock heating is important  this is
only a very crude approximation of the $T-\rho$ diagram of
hydrodynamical simulation (Theuns \etal 1998, Dav\'e \etal 1999, Croft
\etal 2002).  However, it is mainly the moderate density regions which
are responsible for the Ly$\alpha$ forest.  Eq.~(\ref{statesim})
should thus be sufficient for our purposes.  The gas temperature
$\hat{T}$ at fixed overdensity is expected  to  evolve with redshift.
Schaye \etal 2000 and  Theuns \etal 2002 claim to have detected a peak
at $z=3$ with decreasing temperature towards lower and higher
redshifts.  Note that we have ignored a possible  redshift dependence
and have assumed  $\hat{T}=15000\kel\simeq 10^{4.2}\kel$  independent
of redshift.

\subsubsection{The effect of wind bubbles} 

We have  adopted  a very simple model for the effect of wind bubbles.
We  simply assume that shocks produce  long-lived {\it fully ionised}
spherical bubbles around galaxies (see appendix A for a discussion of
the physical properties of expanding wind bubbles). We  assume   that
the neutral hydrogen \hi\ density  is  zero in a spherical region  of
radius $\rw$ centered on the  DM haloes which we have chosen to
identify as LBGS.  Note that this simple model has also been discussed
in Croft \etal (2002), Kollmeier et al. 2003 (2003a,2003b) and
Weinberg et al. (2003).  In the simulation, most of the haloes lie
along filaments, or at the intersection of filaments, and the gas
distribution around the haloes is often anisotropic on the scale of
galactic winds (e.g. Croft \etal 2002, Springel \& Hernquist
2003). Theuns \etal (2002) have also shown that winds propagate more
easily into the low density IGM than in the filaments, giving rise to
highly ionised regions which are generally not spherical.  It should
thus be kept in mind that our assumption  of a spherical wind bubble
will only be  a reasonable approximation for strong isotropic
winds. Furthermore  the bubbles do not necessarily live longer than a
Hubble time.   A discussion of the physical properties of shocks in
the IGM can be found in Appendix~\ref{cooltimescale}.

\begin{figure*}
\resizebox{0.45\textwidth}{!}{\includegraphics{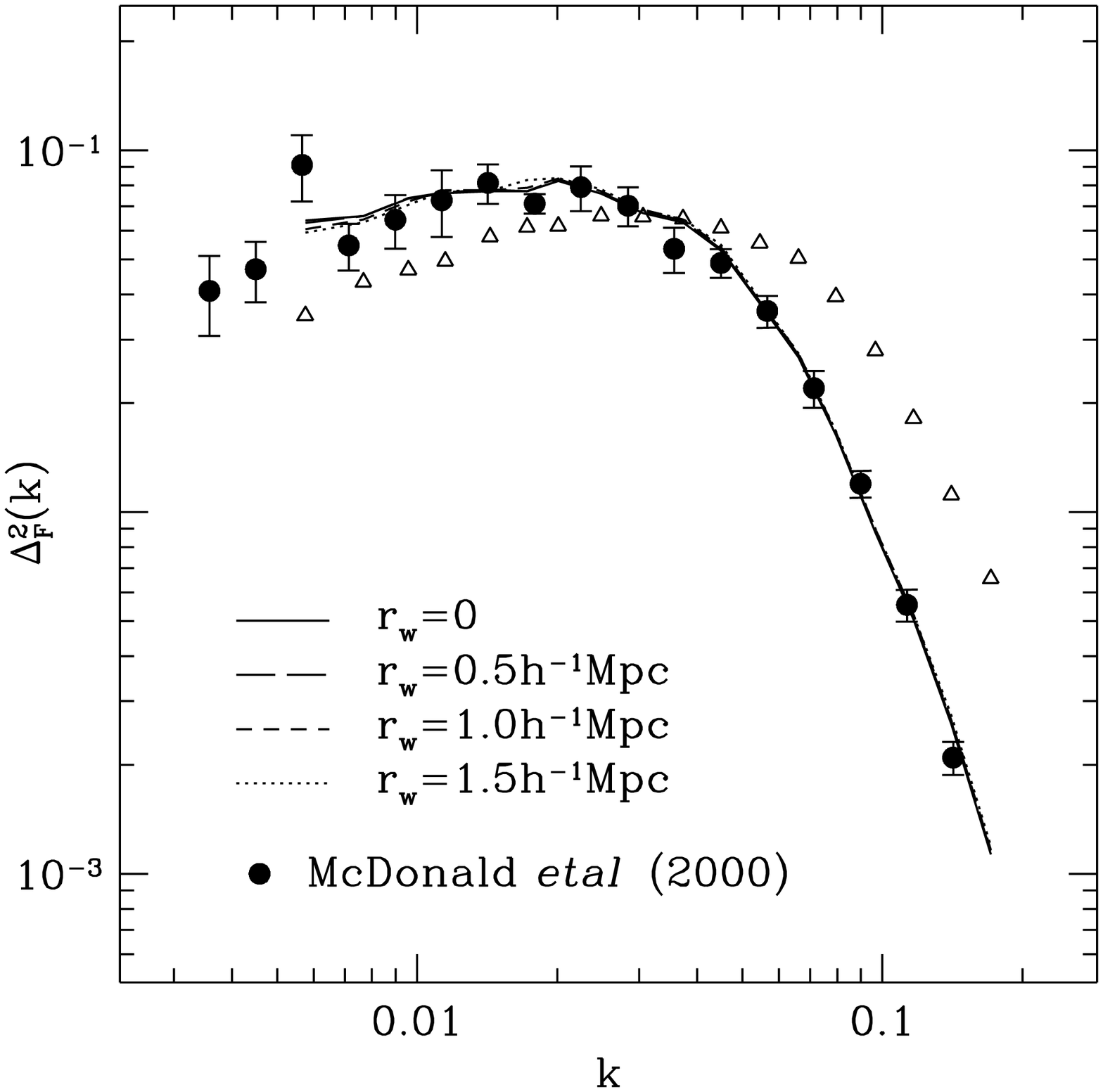}}
\resizebox{0.45\textwidth}{!}{\includegraphics{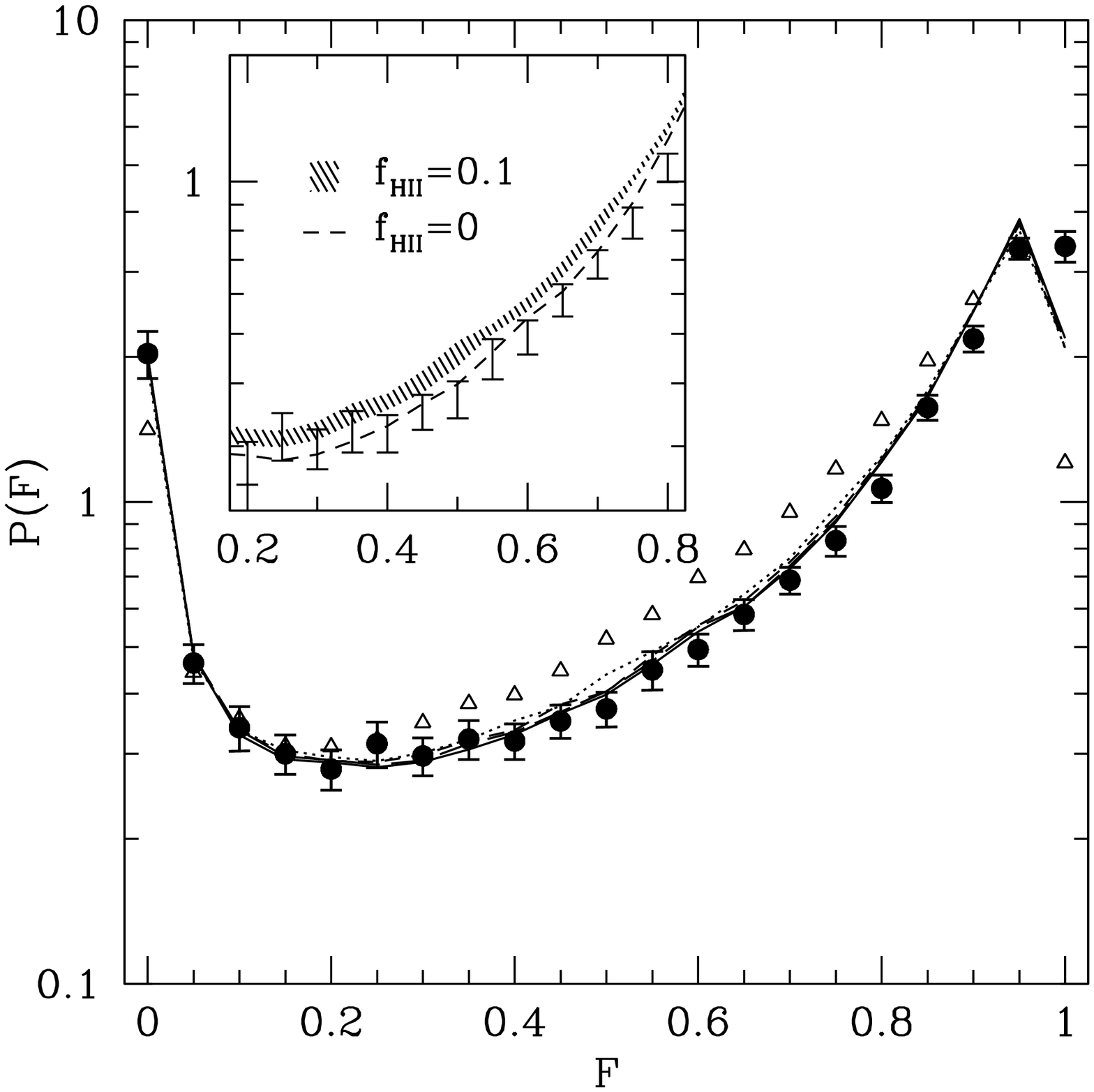}}
\caption{The flux power spectrum (left panel) and PDF (right panel) at
$z=3$ with wind bubbles  of radius $\rw=0$ (solid line), $\rw=0.5$
(long-dashed line), $\rw=1.0$  (short-dashed line) and  $\rw=1.5\hmpc$
(dotted line). The results are compared to the observations of
McDonald  \etal (2000), shown as filled symbols.  The triangles are
the various volume  averaged statistics in real space. The small plot
in the right panel shows the PDF for filling factors of zero and ten
percent (cf. text). The  wavenumber $k$ is in unit of $\kms$.}
\label{fig2}
\end{figure*}

\subsubsection{Properties of the synthetic spectra} 

Once we have  smoothed the dark matter density $\rhof$, the flux
distribution depends only on $\cal{A}$ and $\alpha$ and the  size
distribution of wind bubbles.    We have assumed  $\alpha=1.6$ and
constrain $\cal{A}$ from a random sample of 500 lines of sight of
comoving length $L=30\hmpc$ by demanding that  the mean flux $\la
F\ra$ is equal to the observed value $\la F\ra=0.67$ (Rauch \etal
1997, McDonald \etal 2000). The effect of varying  the assumed size
distribution of wind bubbles are discussed in the following
sections. The  spectral resolution is $\Delta v\simeq 6\kms$, somewhat
smaller than in the ASSP data, where it is $\Delta  v\lsim 10\kms$.
Note that the mean inter-particle distance in the simulation is $\sim
0.2\hmpc$. This  is sufficient to resolve $\bar F$ on a scale $r\sim
0.25\hmpc\sim 27\kms$ as in the observations,  but is somewhat too
low  to resolve all the features  of the Ly$\alpha$ absorption lines
(e.g. Theuns \etal 1998, Bryan \etal 1999). We  also add  noise with
signal-to-noise S/N=50 per  pixel of width $\Delta\pi=2.5\kms$.

\subsection{Volume averaged statistics: The flux PDF and the flux power 
spectrum}
\label{pdf}

We will first investigate some basic flux statistics  of our synthetic
spectra to check if they are a  reasonable representation of observed
spectra and to study the effect of the wind bubbles.   We begin with
the flux probability distribution function (PDF)   defined as the
fraction of the volume of the IGM which has  a flux value within a
certain range.  The right  panel of Fig.~\ref{fig2} compares the flux
PDF at $z=3$ for wind bubbles with different radii $\rw$. The filled
circles show the observed PDF from McDonald \etal 2000, while the
solid, long-dashed, short-dashed and dotted curves correspond to
$\rw=0$ (i.e., no wind), 0.5, 1 and $1.5\hmpc$, respectively.
Following McDonald \etal (2000) we use 21 bins of width  $\Delta
F=0.05$ to compute the flux PDF.  The PDF of spectra without wind
bubbles (solid curve) agrees very well with the PDF of the observed
spectra (filled circles).  Including wind bubbles with $\rw=0.5\hmpc$
has no significant effect on the flux PDF, and  even when $\rw$ is as
large as $1.5\hmpc$ there is only a slight discrepancy  with the flux
PDF of the observed spectra in saturated regions with   $F\sim 0$.  The
small effect of the wind bubbles is due to their small  volume filling
factor $f_{\rm HII}$, which is about 5\% for the model with
$\rw=1.5\hmpc$.  This is consistent with the findings of  Theuns \etal
(2002) and Weinberg et al. (2003).

\begin{figure*}
\resizebox{1\textwidth}{!}{\includegraphics{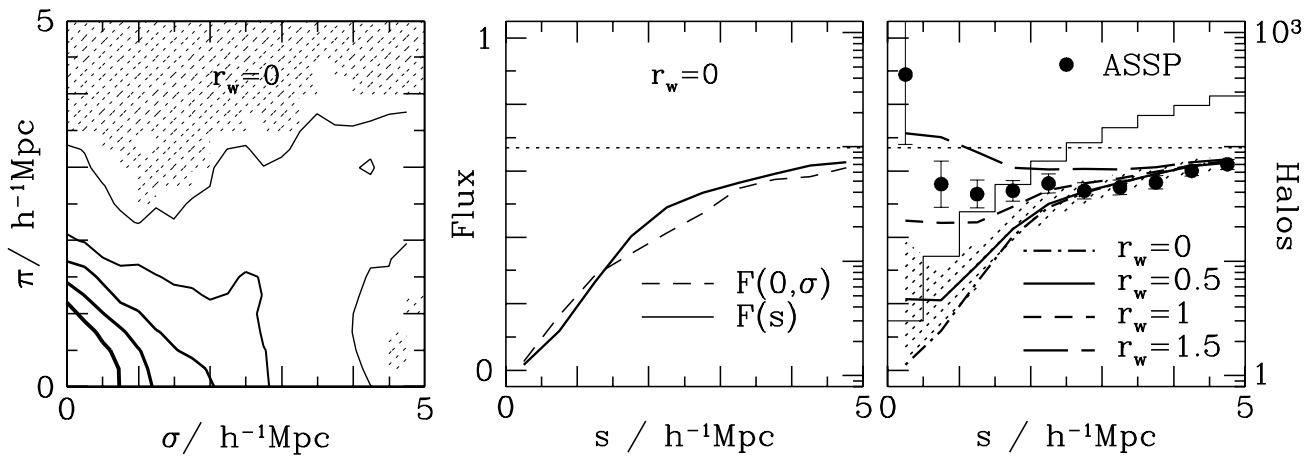}}
\caption{{\it Left panel}~: Contours of the 2D conditional flux ${\bar
F}(\pi,\sigma)$ around the nearest LBG-size halo in the simulation at
$z=3$ (cf. text). The (comoving) bin size is  $\Delta
s=0.25\hmpc$. The shaded pixels have a flux within 0.05 of the mean
flux level, $\la F\ra=0.67$. {\it Middle panel}~:  The mean
transmitted flux ${\bar F}(s)$ at distance  $|{\bf
s}|=\sqrt{\pi^2+\sigma^2}$, compared to ${\bar F}(0,\sigma)$.  The
dotted curve shows the mean flux level. {\it Right panel}~: The mean
transmitted flux, when LBG-size haloes were surrounded by wind bubbles
with radius  $\rw=0.5$, $\rw=1\hmpc$ and $\rw=1.5\hmpc$, respectively,
as indicated in the figure. The Adelberger \etal (2003) measurements
are shown as filled symbols. The shaded area is the 1$\sigma$ error
for the model $\rw=0.5\hmpc$, and was computed from $N(<s)$ shown as
an histogram (right axis).}
\label{fig3}
\end{figure*}

We can  place an upper bound on the volume filling factor, $f_{\rm
HII}$, of bubbles by demanding a good match between the simulated and
observed  PDFs. We have computed the flux PDF for several bubbles
radii for which  $f_{\rm HII}$  is in the  range 0.01-0.5. To account
for the  uncertainty in the mass of LBGs, we performed three
computations for each  $f_{\rm HII}$~: we selected DM haloes with mass
$M$ larger than $5\times 10^{10}$, $10^{11}$ and $5\times
10^{11}\mdh$, and adjusted $\rw$ which gives the desired $f_{\rm
HII}$.  We find that independent of halo mass, a  filling factor
$f_{\rm HII}\bsim 0.1$ yields  a poor match to the observations,
especially in the range  $0.2\lsim F\lsim 0.8$, as shown in the small
plot inside the  right panel of Fig.~\ref{fig2}.

In the left panel of  Fig.~\ref{fig2}  we show the dimensionless
one-dimensional flux power $\Delta^2_{\rm F}(k)$ as a function of $k$
(in $\kms$), defined as
\begin{equation}
\Delta^2_{\rm F}(k)=\frac{L}{\pi}k \,|\delta_k|^2\; ,
\label{ps}
\end{equation} 
where $\delta_k$ is the Fourier Transform of the flux contrast $(F/\la
F\ra-1)$.  The flux power spectrum was calculated  with a  FFT routine
from 500 synthetic spectra.  Note that  our  simulation box and
therefore the spectra are  not periodic which may lead to small
artifacts at small scales.  The  flux power of our synthetic spectra
and that of the  observed spectra obtained by  McDonald \etal (2000)
also agree  very well. Bubbles of size as large as $\rw=1\hmpc$ have
again   a very small effect due to their small volume filling  factor
(e.g. Croft \etal 2002, Weinberg et al. 2003).

To demonstrate the role of peculiar velocities, the triangles in
Fig.~\ref{fig2} show  the flux PDF and the power spectrum in real
space for $\rw=0$ obtained  by setting the peculiar velocities and
thermal motions  to zero.  The real space flux PDF and power spectrum
significantly deviate  from the observations.

\section{Probing correlations between flux and haloes/LBGs}
\label{xihf}

\subsection{Statistical measures of  the flux-halo correlation} 

There is a variety of statistical measures which can describe the
correlation between the flux along a LOS and nearby galaxies. 
Here we will focus on the conditional flux function, $\bar F$. 
$\bar F$ is the mean flux at real-space distance $|{\bf s}|=s$ to 
the next halo,  
\begin{equation}
\bar F({\bf s}) = \frac{1}{N_h}\sum F({\bf s} |{\rm halo}) \; ,
\label{fcond}
\end{equation}
where $N_h$ is the total number of haloes, $F(s|{\rm halo})$ is the
value of the flux at a distance $s$ from the nearest halo.   
Note that we use the position of the haloes which we identify 
as host haloes of LBGs  as a proxy for the galaxy position. 
Results will thus depend on how this identification is done and 
we will later explore several possibilities. 
The summation is performed over a representative sample of  haloes.   
The function $\bar F$ can be expressed in terms of the unconditional
two-point correlation $\xihf$   as
\begin{equation}
\bar F({\bf s}) =\la F\ra\left[1+\xihf({\bf s})\right] \;.
\label{corr.con}
\end{equation}
Relation  (\ref{corr.con}) also holds in redshift space.  In the
following  $\pi$  and $\sigma$ are the coordinates of ${\bf s}$
parallel and perpendicular to the LOS, respectively.

\subsection{Calculating the mean flux as function of the distance to
the nearest halo}
\label{fluxobs}

For the numerical simulation we know the full 3-dimensional  flux
field. We could therefore estimate $\bar F(s)$ from  the transverse
correlation ${\bar F}(\pi=0,\sigma)$.  However, real observation will
sample the flux field $\bar F({\bf s})$ more sparsely. ASSP  have
therefore estimated $\bar F(s)$ from the averaged  2D halo flux
correlation function $\xihf$ using eq.~(\ref{corr.con}) in order to
maximize the signal.  We proceed as ASSP do and calculate $\bar F (s)$
as follows. We first pick a LOS direction. We then  produce  an
ensemble of spectra along this direction with random offsets
perpendicular to the chosen direction. For each pair of halo/spectrum
we determine the flux ${\bar F}(\pi,\sigma)$,  the distance $\sigma$
at the point of closest approach, and the distance $\pi$ between the
halo and a given pixel of the spectrum.  This process is repeated for
an ensemble of randomly chosen LOS directions.  We then obtain  $\bar
F(s)$ by annular averaging   ${\bar F}(\pi,\sigma)$, and binning   in
bins of comoving size $\Delta s$. We further average  over all haloes
and LOS.  In the left panel of Fig.~\ref{fig3} we plot the full 2D
distribution of the average flux $\bar F(\pi,\sigma)$   around haloes
assumed to host LBGs. The haloes were assumed  to have no  galactic
wind bubbles. The bin size is $0.25\hmpc$, twice smaller than that we
use for $\bar F$.   The shaded area shows bins for which $\bar F$ is
within 0.05 from the mean flux $\la F\ra$.  Contours are for flux
levels increasing  from   0.2 to 0.6 with decreasing line width. The
contours are compressed  along the line of sight as a result of the
peculiar and thermal motions of the gas.  At small separation the
transmissivity is much  smaller  than the  mean due to the increased
density near haloes.

\begin{figure*}
\resizebox{1\textwidth}{!}{\includegraphics{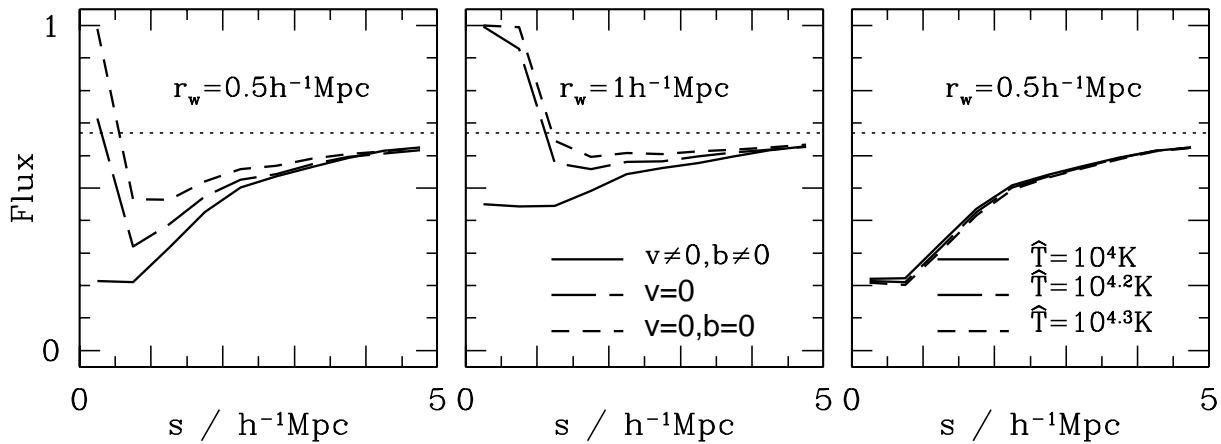}}
\caption{ {\it Left and Middle panels}~: The transmitted flux in
redshift space (solid), with gas and halo velocities set to zero but
with non-zero Doppler parameter $b$ (long-dashed),  and with
velocities and $b$ set to zero (short-dashed).   {\it Panel to the
right}~: The transmitted  flux as a function of the  mean IGM
temperature $\hat{T}$ as indicated on the figure.  In each panel the
horizontal dotted line shows the mean flux level, $\la F\ra=0.67$.}
\label{fig4}
\end{figure*}

Redshift distortions cause $\bar F(s)$ (as computed in ASSP) to be
different from the transverse correlation $\bar F(\pi=0,\sigma)$.
This can be seen in the middle panel of Fig.~\ref{fig3}. The effect is
important for $s\sim 1-3\hmpc$. Note that previous work compared also
the ASSP measurements with $\bar F(s)$ obtained  from numerical
simulations (e.g. Croft \etal 2002, Kollmeier \etal  2003, Bruscoli
\etal 2003).

\subsection{Estimating errors due to sparse sampling} 
\label{errors}

The sample of ASSP  contains 431 galaxies  in 7 fields containing a
QSO (one field contains two QSOs).  At large scales  ($>$ a few Mpc)
the number of galaxy LOS pairs is reasonably large. However
at small  scales the number of available galaxies becomes very small.
There are {\it e.g.} only 3 galaxies within a transverse separation of
$0.5\hmpc$ to the line-of-sight to a QSO.   Obviously Poisson errors
will then be large  and estimating these errors is thus important.  In
the following we use the term {\it cosmic  variance} to  describe the
error resulting from the finite number of  galaxies and QSO spectra.
At small scales cosmic  variance is the dominant source of
error. Other sources of errors also affect the calculation of $\bar
F$, mainly the uncertainty of  the mean flux $\la F\ra$, and the
redshift position of the galaxies. These errors will be discussed  in
Section~\S\ref{e:role}.

About one LOS from a bright QSO is expected  to pass through a region
of angular dimensions comparable to that of   our simulation box at
$z=3$. However, the correlation length of the flux is around a few
$\hmpc$. There will thus be about a hundred LOS  through the
simulation box  which represent independent flux measurements.  We
therefore estimate the cosmic variance error as follows  from the
simulation.  For each separation bin of width $\Delta s=0.5\hmpc$  we
perform  Monte-Carlo  realizations of $\bar F(s)$, where $\bar F(s)$
is averaged  over the  halo/spectrum pairs  lying in the given bin. The
cosmic variance error is then the $1\sigma$ scatter around $\bar
F(s)$.    The error of $\bar F(s)$ depends on  the total number of
galaxies/haloes  with separation $s$. A sample of about 40 LOS
through our simulation  box gives a cumulative halo function $N(<s)$
close to the cumulative galaxy function $N_{\rm LBG}$ of the ASSP
sample for separation $s \leq 1\hmpc$.   Note also that the number of
haloes with  $M>M_{\rm min}$ in the simulation is only about 110.  This
means that we severely oversample the halo catalog when we estimate
the cosmic variance on large scales. However, at  the small scales in
which we are mainly interested this  should have little effect.

\section{The halo flux correlation of observed and simulated 
spectra}
\label{fluxsim}

\subsection{Wind bubbles and the halo flux correlation} 

We first explore the shape of $\bar F$ assuming that all wind bubbles
have the same radius, $\rw$, independent of the halo mass that they
surround.  In the right panel of Fig.~\ref{fig3} we plot $\bar F$ for
haloes with mass $M\geq M_{\rm min}$. We show curves for $\rw=0.5$, 1
and $1.5\hmpc$ as indicated in the figure. The corresponding filling
factors are $f_{\rm HII}\sim0.2$\%, $\sim 1.6$\% and $\sim 5.3$\%
respectively, small enough to ensure that the flux PDF, the power
spectrum and the line distribution of the \op\ forest are not
affected. The dashed-dotted curve corresponds to $\bar F$ in the
absence of outflows. The filled symbols show the observed correlation
as given in ASSP. The shaded area shows the cosmic variance error, and
is calculated (cf. Section~\S\ref{errors}) from the cumulative halo
distribution $N(<s)$  also  shown as a histogram in the right panel
of Fig.~\ref{fig3} (right axis).  The cosmic variance error is only
shown  for $\rw=0.5\hmpc$.  

\begin{figure*} 
\centering
\resizebox{1.0\textwidth}{!}{\includegraphics{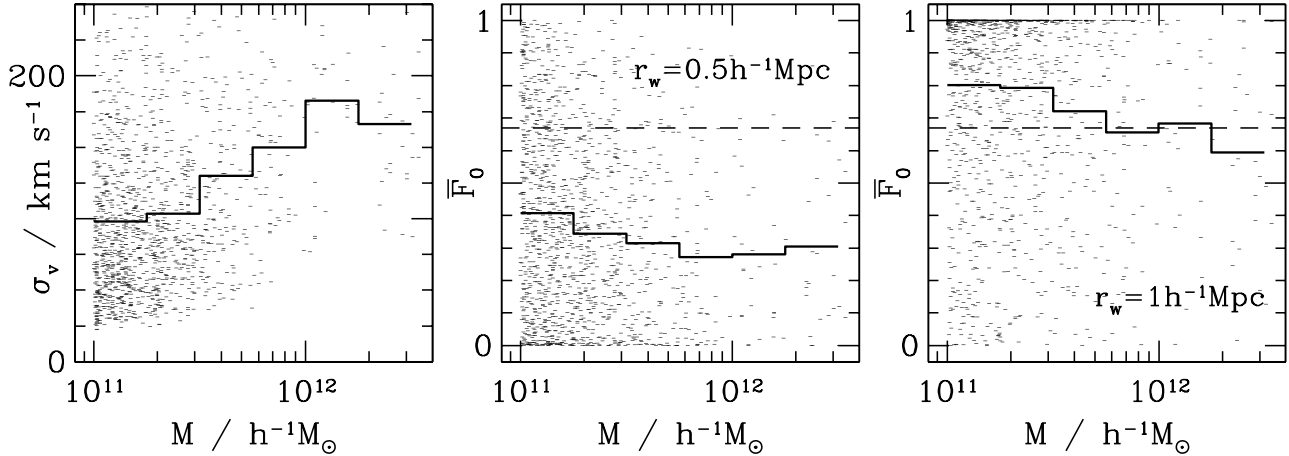}}
\caption{{\it left panel}~: The velocity dispersion along the line of
sight $\sigma_v$ near haloes of mass $M\geq 10^{11}\mdh$.  The solid
line is $\sigma_v$ averaged over bins of size $\Delta\log
M=0.25$. {\it Middle and right panels}~: the transmitted flux at
separation $s\leq 0.5\hmpc$, ${\bar F}_0$, as a function of $M$. The
horizontal dashed  line is $\la F\ra=0.67$.}
\label{fig5}
\end{figure*}

\begin{figure*} 
\resizebox{0.45\textwidth}{!}{\includegraphics{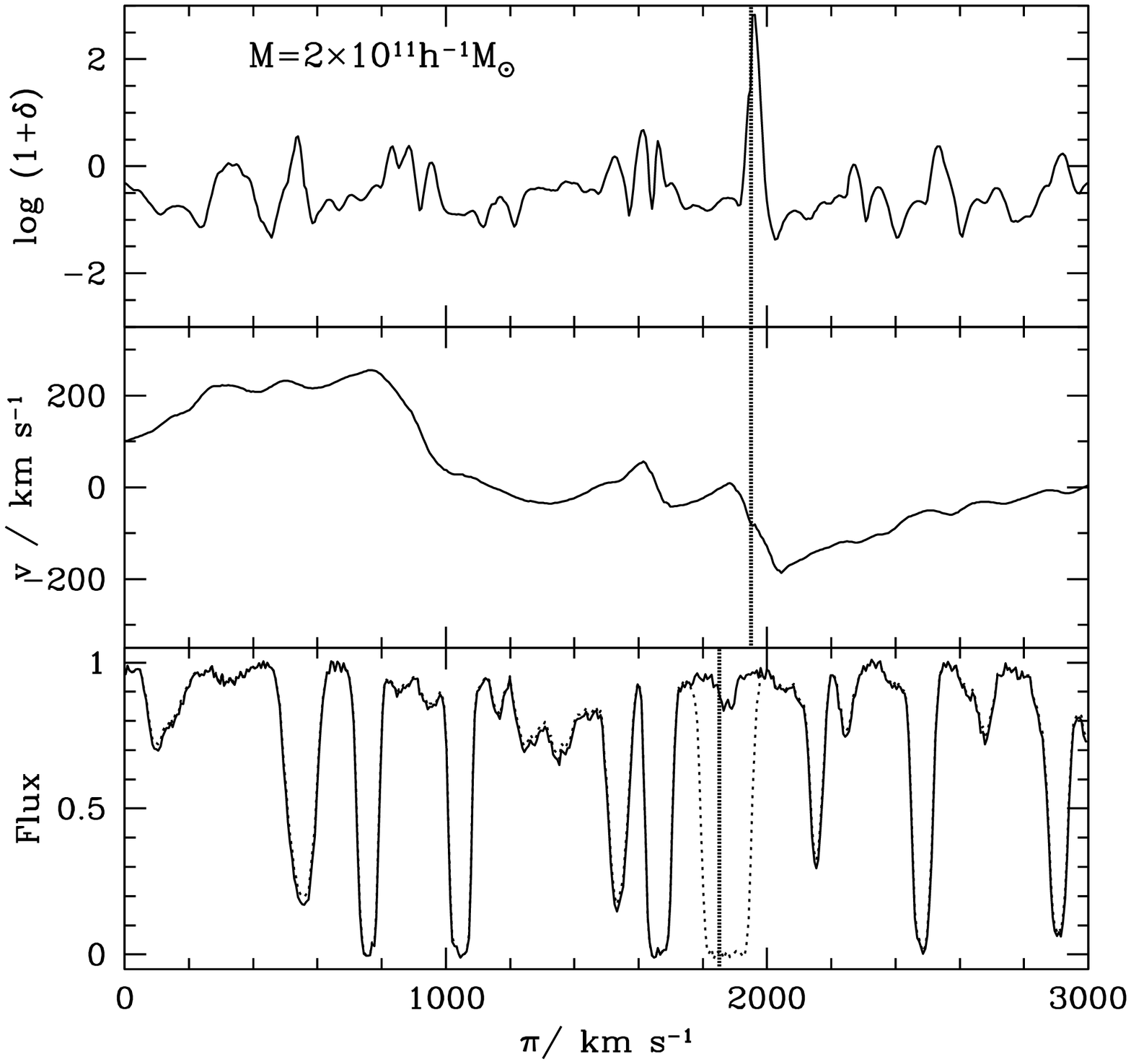}}
\resizebox{0.45\textwidth}{!}{\includegraphics{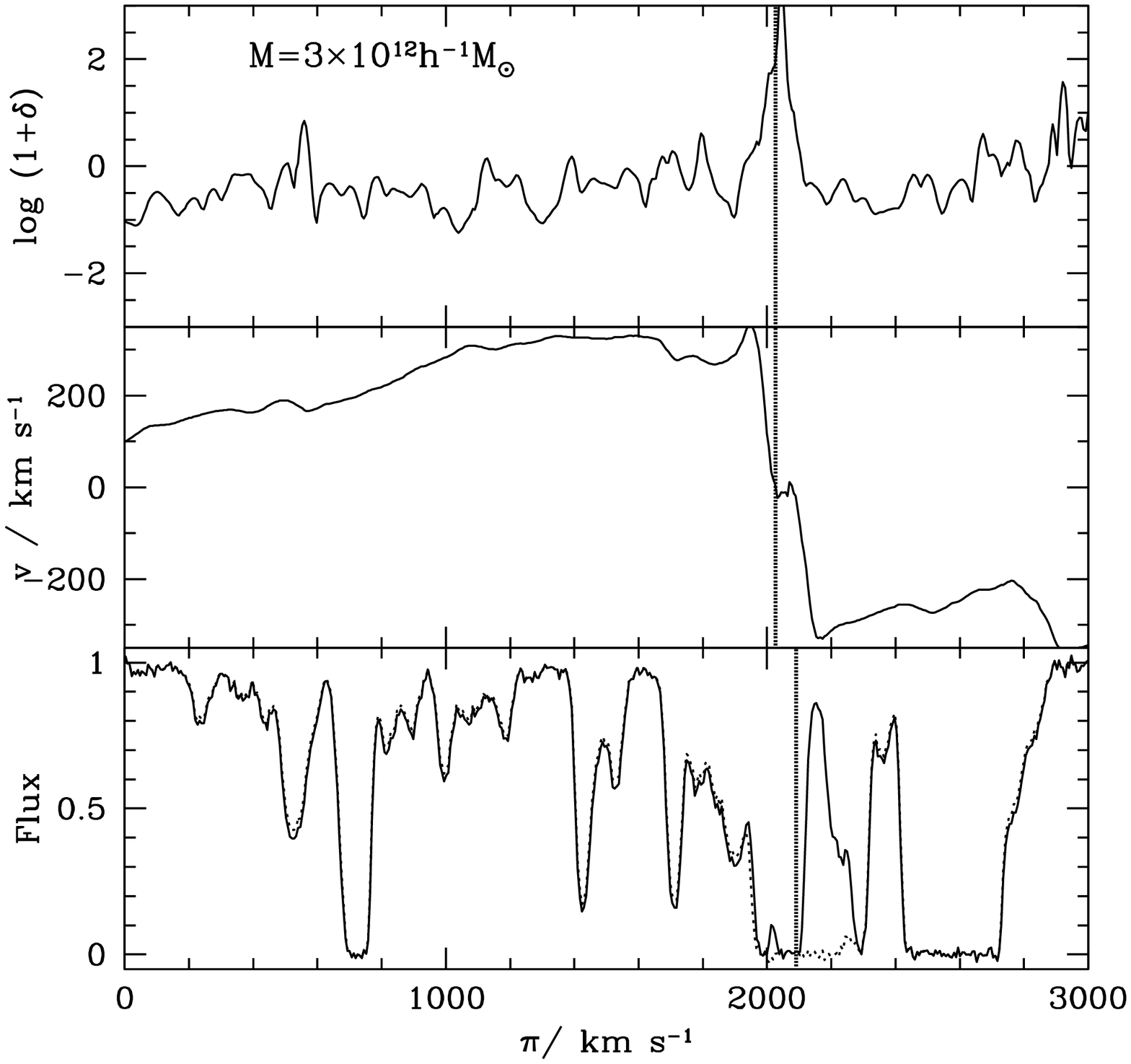}} 
\caption{One-dimensional gas density, velocity (real space) and
transmitted flux (redshift space) along two lines of sight extracted
from simulation at $z=3$, running trough the centers of haloes with
mass $M=2\times 10^{11}$ (left panel) and $M=3\times 10^{12}\mdh$
(right panel). The halo positions  along the line of sight are shown
by vertical lines. The transmitted  flux is computed with winds (solid
line) and without winds (dotted line).  The bubble radius is
$\rw=0.5\hmpc$.}
\label{fig6}
\end{figure*}

The trend of decreasing mean flux with decreasing distance to the next
galaxy in the observations is well reproduced by the numerical
simulations. At small separation this trend is inverted if the wind
bubbles are included. As Kollmeier et al. (2003a, 2003b)  we find that 
the strong increase of  the mean flux level at  $s \bsim 0.5\hmpc$, 
is only reproduced for bubbles with size as large as $r_w\bsim 1.5
\hmpc$.  We thus confirm the difficulty  of  
other authors to  reproduce the  mean flux   at small separation
unless rather large bubbles are included.  Bubbles of this size are,
however,  difficult to  explain even with  very efficient galactic
superwinds (Croft et al. 2002).

There are thus two possibilities: the  simple  model  of spherical
wind bubbles of constant size around  massive haloes does not describe
the effect of winds  of  observed galaxies properly or the observed
strong increase of the flux level in the ASSP is a statistical
fluke. We will explore both  possibilities in the following sections
in more  detail.

\subsection{The effect of  peculiar motions and thermal broadening}
\label{v:role}

\begin{figure*} 
\resizebox{1.0\textwidth}{!}{\includegraphics{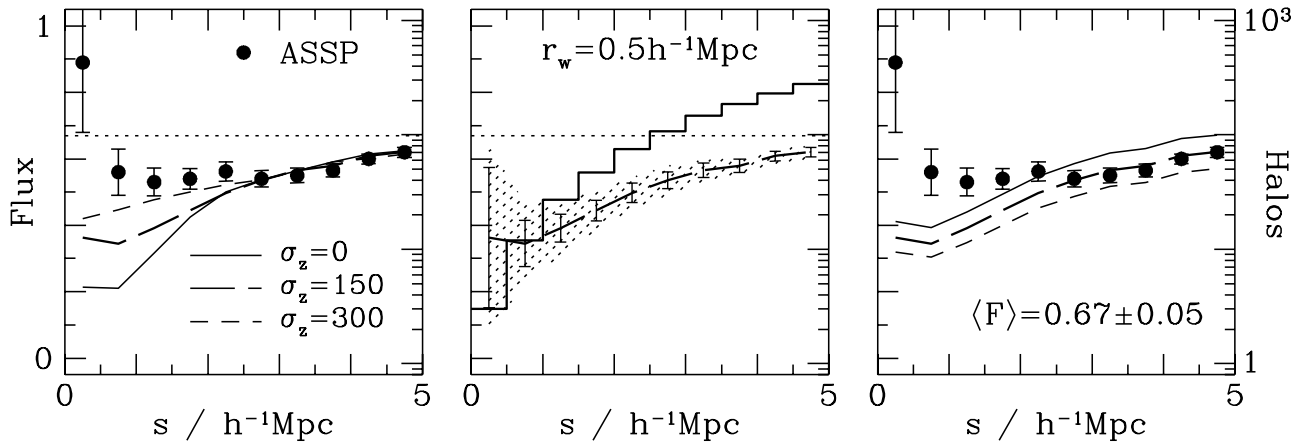}}
\caption{{\it Left panel}~:  The transmitted flux $\bar F$ for bubbles
of size $\rw=0.5\hmpc$ after adding a Gaussian deviate of variance
$\sigma_z=0$, $\sigma_z=150$ and $300\kms$ to the LOS halo position.
{\it Middle panel }~: The cumulative halo function (histogram) and the
1$\sigma$ error (shaded area) for the model with $\sigma_z=150\kms$.
{\it Right panel}~: The transmitted flux with a mean flux level $\la
F\ra$=0.62, 0.67 and 0.72. The filled symbols are the measurements by
Adelberger \etal.}
\label{fig7}
\end{figure*}

Kollmeier \etal (2002) have pointed out that the peculiar velocities
of matter infalling onto the galaxy/DM halo expected from hierarchical
structure formation should have a strong influence on $\bar F(s)$  at
small separations. In Fig.~\ref{fig4} we  investigate  the effect of
coherent small-scale peculiar velocities  and  thermal velocities. In
the left and middle panel  we show $\bar F(s)$  with and without
peculiar and thermal velocities  for wind bubbles of size  $r_w=0.5
\hmpc$ and  $r_w=1\hmpc$, respectively. Without peculiar velocity the
increase of the mean flux at small separation is much more
pronounced. This is easy to understand. In redshift space  the
velocity shear  of the infalling surrounding  material  fills in the
cavity which is  caused by the fully ionised wind bubble in real
space. For the smaller bubble size such an effect is  also visible
for the thermal velocities alone.  Neutral hydrogen \hi\ lying at the
boundary of the ionised bubbles leads  to significant \op absorption
at the redshift halo position when the ratio $b/\rw$ is large enough.
The peculiar velocity will, however, always dominate over the effect
of thermal velocities. This can be seen in the right panel where we
have scaled the temperature of the gas up and down.  This has  very
little effect on $\bar F(s)$  calculated with peculiar velocities even
for the smaller bubble size.
We have also computed $\bar F(s)$ for various values of the  density
threshold in eq.~(\ref{statesim}). We found only small  differences
for $10\lsim \delta\rho/\rho\lsim 100$.  Such overdensities occur
typically within $\lsim 0.2\hmpc$ (comoving)  from the halo
(Kollmeier \etal  2003) which is smaller than the bubble radii
considered here.  Note that the good agreement of the flux PDF of the
synthetic spectra  with the McDonald \etal (2000) data found in
section 3  disappeared if peculiar velocities were set to zero. This
indicates that our simulations have similar peculiar velocities  as
the gas responsible  for the  observed absorption -- at least in a
volume-averaged sense.

In the left panel of Fig.~\ref{fig5} the black dots show the velocity
dispersion for haloes of mass $M$.  For each DM halo  the velocity
shear  $\sigma_v$ along the line of sight was calculated along  three
perpendicular  LOS directions for  an interval $\Delta\pi=2\hmpc$
centered on the halo position.  The histogram shows the mean velocity
dispersion  averaged over bins with $\Delta\log M=0.25$.   The
velocity shear  clearly increases with halo mass.  This trend is
present for a range of  $\Delta\pi$, $1\hmpc\lsim\Delta\pi\lsim
2\hmpc$.    The middle and right panels of Fig.~\ref{fig5} show the
transmitted flux at separation $s\leq 0.5\hmpc$,  ${\bar
F}_0=F(s<0.5)$, as a function of the halo  mass $M$ assuming  that all
haloes with $M\geq 10^{11}\mdh$ are surrounded with bubbles  of radius
$\rw=0.5$ and $\rw=1\hmpc$, respectively. As expected ${\bar F}_0$ is
significantly larger for $\rw=1\hmpc$.  The dependence on  halo mass
appears to be stronger for $\rw=1\hmpc$ than for $\rw=0.5\hmpc$. In
the former case,  ${\bar F}_0$ decreases with increasing $M$ by about
$\Delta {\bar F}_0=0.20$ from $M=10^{11}$ to $M=10^{12}\mdh$. This is
consistent with the trend in the left panel which predicts that low
mass haloes satisfy the condition $\rw\bsim (1+z)\sigma_v/H(z)$  more
often than massive haloes as a result of the $\sigma_v-M$
dependence. For smaller bubbles, $\rw\lsim 0.5\hmpc$, this appears to
be rarely the case, independent of the mass of the halo.

To investigate the effect of the  velocity field near haloes further
we show the real space density and velocity field of the gas,  as well
as the absorption spectrum  along two  lines of sight passing haloes
with $M=2\times 10^{11}$ (left panel) and $3\times 10^{12}\mdh$ (right
panel) in Fig. ~\ref{fig6}. Solid and dotted line are for the case
with and without wind bubbles, respectively. The radius of the bubbles is
$\rw=0.5\hmpc$ for both haloes.  The difference between the two cases
is striking . For the smaller mass halo  in the left panel the flux
level at the position of the halo is  $\bar F\sim 0.9$  if the bubble
is included  whereas $\bar F$ saturates at the redshift space position
of the more massive halo in the right panel. Since the density fields
near the haloes in both panels show similar features, the difference in
the flux values is mainly due to the difference in the velocity fields
near the two haloes.  The velocity shear in the less massive halo (left
panel) is quite modest. On the contrary, the velocity field near the
massive halo (right panel) has a large infall on a physical scale of a
few hundred $\kms$, which moves  some of the moderate density peaks
close to the redshift space position of the halo.  This effect occurs
for a large fraction of haloes, and thereby reduces the overall
expected $\bar F$ at  small scales by a few tens of percent.

\subsection{The role of galaxy redshift error, LOS smoothing, and 
flux normalization}
\label{e:role}

There is a number of other uncertainties which have to be taken into
account when comparing the observed $F(s)$ with that calculated  from
the numerical simulation. Most important are the uncertain  redshifts
of the LBGs.  There are large systematic shifts  of several hundred
$\kms$ between the redshift of nebular emission lines, \op emission
and interstellar absorption lines which make the assignment of a
redshift somewhat ambiguous.  ASSP give $\Delta z\geq 0.002$  or about
$150\kms$ as a typical error.  We know however relatively little about
the dynamical state of the  responsible gas and there may well be
larger systematic errors.  The left panel of Fig.~\ref{fig7} shows the
effect of adding a  Gaussian distributed error to the redshift of the
DM haloes.    The solid curve shows the same model as the solid curve
in the right panel of Fig.~\ref{fig3} (``massive'' haloes,
$\rw=0.5\hmpc$).  Long-dashed and short-dashed curves are for redshift
errors of  $\sigma_z=150\kms$ and $300\kms$, respectively.   As
expected introducing redshift errors smoothes out the large depression
of the flux level at distances of  up to $\sim 2\hmpc$ to the
line-of-sight.  This substantially improves the agreement with  the
ASSP measurements on these scales.

In the middle panel of Fig.~\ref{fig7} the shaded area shows the
1$\sigma$ error of the simulated $F(s)$ for our ensemble of LBG-size
haloes with mass $M\geq 5\times 10^{11}\mdh$. The errorbars show
the errors quoted by ASSP. To make the comparison easier the  errors
are plotted for the  model with  $\sigma_z=150\kms$.  The solid
histogram shows the cumulative halo function $N(<s)$ (right axis) as a
function of separation.  At $s\leq 1\hmpc$ the cumulative halo
function $N(<s)$  is close to the cumulative galaxy function $N_{\rm
LBG}$ of the ASSP  sample, with $N(s<0.5)=3$ and $N(s<1)=11$ (resp. 3
and 12 in the ASSP sample).    Our errors are generally larger than
that of ASSP,  at  $s\leq 0.5\hmpc$ by about 30 percent.  At larger
scales the difference is smaller but note that there our sample is
about 20\% larger than that of ASSP.  Note also that the errors
decrease faster than those of the transverse correlation $F(0,\sigma)$.

The mean flux level of QSO spectra  varies significantly from spectrum
to spectrum (10-20 percent). We thus also investigated  how $\bar F$
depends on the value of the mean flux $\la F\ra$. In the right panel
of Fig.~\ref{fig7} we show the effect of varying $\la F\ra$.  The
long-dashed curve is our fiducial ``massive'' haloes model, which has
$\rw=0.5\hmpc$ and $\sigma_z=150\kms$.   The upper (solid) and lower
(short-dashed) curves show $\la F\ra=0.72$ and 0.62  respectively.  A
larger (lower) mean transmitted flux $\la F\ra$ raises (lowers) $\bar
F(s)$ almost linearly by the same amount $\Delta F=\pm 0.05$ at all
separations. $\la F\ra=0.67$ matches the large scale ASSP measurements
best.

\subsection{The effect of varying the bubble model}
\label{m:role}

\subsubsection{$\bar F$ with halo-mass dependent bubble size}

\begin{figure} 
\mbox{\psfig{figure=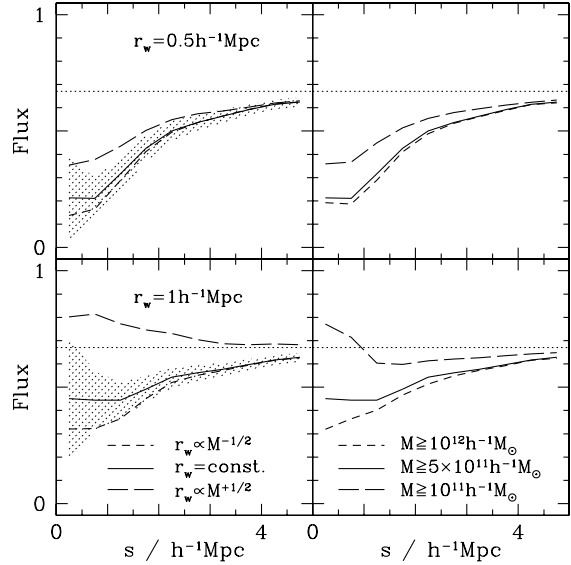,height=3.2in,width=3.2in}}
\caption{{\it Panels to the left}~: $\bar F(s)$ for a mass-dependent
$\rw$ as indicated in the figure.  Panels on the top row have
$\rw=0.5\hmpc$, whereas panels on the bottom row have $\rw=1\hmpc$.
The horizontal dotted line shows the mean flux, $\la F\ra=0.67$.  {\it
Panels to the right}~: the mean transmitted  flux $\bar F(s)$ for
three values of the halo mass limit  $M_{\rm min}$.  The shaded area
is the 1$\sigma$ error, attached to the model $\rw$=const.}
\label{fig8}
\end{figure}

So far we have considered bubbles with sizes that are independent of
mass.  Unfortunately, it  is far from  obvious what scaling with mass
is appropriate. Obviously, on average more stars and a larger  total
energy input from stellar feedback are expected in the more massive
haloes. However, simple binding energy arguments show that winds
should escape more easily from low mass haloes (Larson 1974,  Dekel \&
Silk 1986). This may be properly taken into account by a constant
bubble radius but this is very uncertain. We therefore investigate the
effect of different power-law scalings of the bubble  size with mass.
The left panels of Fig.~\ref{fig8} compare the case of constant bubble
size with those assuming a power-law relation $\rw(M)\propto M^\nu $
with $\nu=+1/2$ and $-1/2$, at the minimal halo mass  $M_{\rm
min}=5\times 10^{11}\mdh$.    The top panels are for  $\rw=0.5$ and
the bottom panels are for   $\rw=1.0$.   For $\nu=+1/2$   $\bar F(s)$
is very sensitive to   the value of $\rw$ and the mean flux at   $s <
1\hmpc$   rises to values as large as   $\bar F =0.8 $ for the larger
of the two bubble sizes improving the agreement with the results of
ASSP.  However, this is at the cost of the good agreement with the
data on larger scales.   It appears thus difficult  to reproduce the
observations of ASSP at small and large scales simultaneously by
changing the mass-radius relation of the bubbles.

\begin{figure*}
\resizebox{0.45\textwidth}{!}{\includegraphics{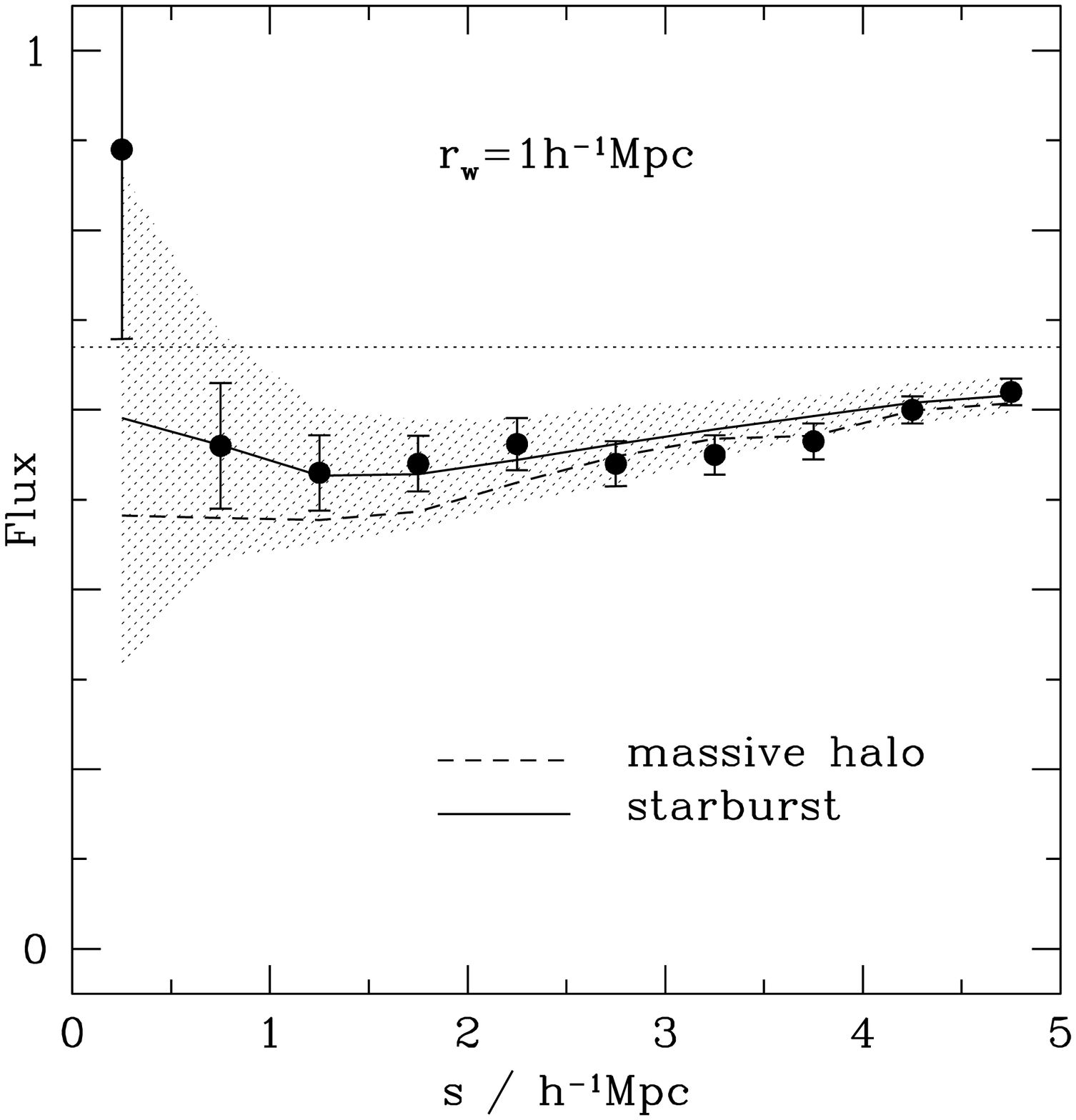}}
\resizebox{0.45\textwidth}{!}{\includegraphics{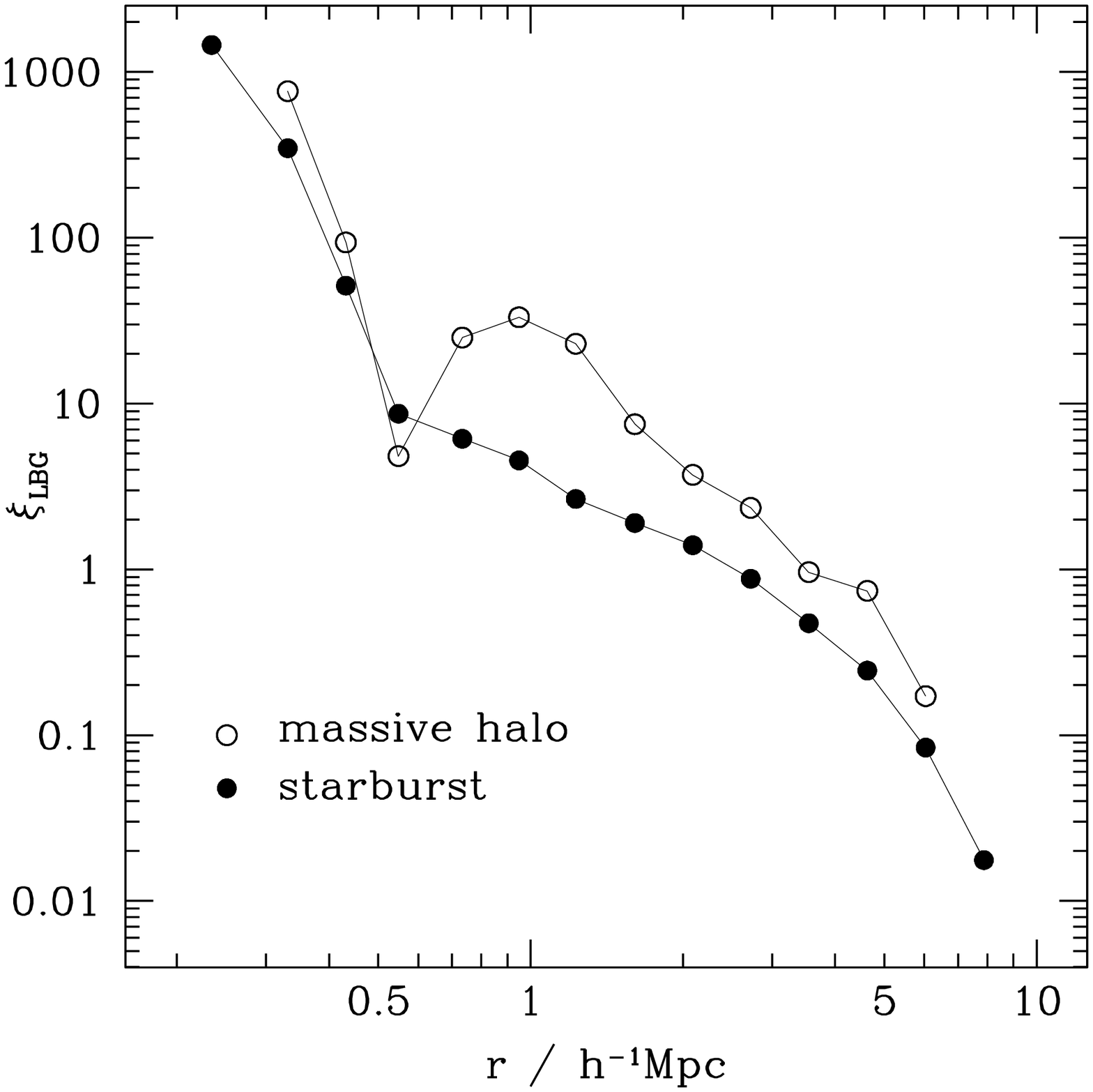}}
\caption{{\it Left panel}~: Comparison between the ``massive-halo''
picture (dashed curve) and the ``starburst'' picture (solid curve).
The shaded area shows the 1$\sigma$ error. A Gaussian deviate of
variance $\sigma_v=150\kms$ was added to the halo positions  {\it
Right panel}~: A comparison between the (real space) two-point
correlation function of LBGs haloes in the ``starburst'' and
``massive-halo'' models. The filled and empty circles show $\xi_{\rm
LBG}$ in the ``starburst'' and ``massive-halo'' models
respectively. On comoving scale $r\sim 0.5-5\hmpc$,  $\xi_{\rm LBG}$
in the ``starburst'' model follows a power-law of index $\gamma=1.6$
and correlation length $r_0=2.5\hmpc$.}
\label{fig9}
\end{figure*}

In the right panels of Fig.\ref{fig8} we investigate the effect  of
changing the minimum mass above which DM haloes are assumed  to be
surrounded by wind bubbles.  Changing $M_{\rm min}$ will change  the
total space density of the halo as given in Table ~\ref{table2}.
There is less absorption if $M_{\rm min}$ decreases, mainly as a
result of the  increasing bubble filling factor .  With a bubble
radius $\rw=1\hmpc$, the filling factor is $f_{\rm HII}\sim 15\%$  and
$\sim 1\%$  for $M\geq 10^{11}$ and $M\geq 10^{12}\mdh$, respectively.

\subsubsection{The starburst model} 

So far we have assumed that LBGs are hosted by the most massive haloes
available and have introduced a minimum mass to match their space
density  with that observed for LBGs. This assumes that LBGS are
long-lived and have the same brightness over a fair fraction of the
Hubble time.  This picture may be in conflict with the lack of the
evolution of the space density of bright LBGs towards very high
redshift which is more easily explained if LBGs are starbursts with a
duty cycle of high star formation rate of less than 10\%.  This would
mean that there is an  about a factor ten or more larger number of  DM
haloes  needed to host these LBGs than in the massive halo picture and
the DM haloes hosting LBGs are less massive on average. As  should be
apparent from the previous section, for a given bubble  size
surrounding a typical LBG halo, this should  reduce the effect  of the
velocity  shear. It is  then easier to reproduce the high mean flux
levels at small separation.  We investigate this alternative
``starburst'' picture here in more detail.  Kolatt \etal (1999)
developed a simple,  {\it ad hoc} procedure to assign LBGs to
colliding haloes identified in their N-body simulation and were able to
reproduce both the number density and clustering properties of
LBGs. Here, we will do something simpler and  randomly select haloes
with mass $M\geq 10^{11}\mdh$ such that their number density matches
the observed LBGs number density, $n_{_{\rm LBG}}\sim 0.004\hhmpc$.
This corresponds  to a duty cycle of ten percent.  Note that only 10\%
of the  haloes with a mass above $10^{11}\mdh$ are haloes as massive as
our ``massive-halo'' picture i.e. have masses $M\geq 10^{12}\mdh$
(cf. Table~\ref{table2}). Most of the LBG-size haloes  have $M\lsim
5\times 10^{11}\mdh$ in the ``starburst'' scenario.  We have produced
120 different Monte-Carlo realizations of such catalogs of
starbursting haloes. In the left  panel of Fig.~\ref{fig9} we compare
$\bar F(s)$ for the ``massive-halo'' model (solid curve) and the
``starburst'' picture (dashed curve). For both we  assume  a bubble
radius $\rw=1\hmpc$ and a redshift error of the halo position of
$\sigma_z=150\kms$.  At $s\lsim 1\hmpc$ $\bar F(s)$ is larger   in the
``starburst'' model by  about 20\%.   We also plot the 1$\sigma$ error
for the starburst  model and  the ASSP  measurements. Our starburst
model with $\rw=1\hmpc$ and a volume filling factor of 2\% appears to
be fully consistent with  the   ASSP measurement. Even at $s\lsim
0.5\hmpc$, where our estimate of the cosmic variance error is 30\% larger
than that of ASSP, $\bar F(s)$ in  the starburst model is only slightly  
more than $1\sigma$ below  the ASSP measurement despite a difference in the
mean flux of $ \Delta \bar F(s) =0.3$.  
Note that for  the quoted volume filling factor  we
did only take into account the bubbles around the haloes that host
LBGs  at a given time. If the bubbles last longer than the LBG phase 
the volume filling factor would be higher by the same factor.

\subsubsection{The clustering of LBG haloes}

As discussed above our procedure to identify the host haloes of LBGs
in the starburst model is rather crude. Unlike Kolatt et al.  (1999)
and Wechsler et al.  (2001) we do not identify ``colliding'' haloes
but choose a fraction of less massive haloes at random. It is thus not
obvious that the host haloes chosen in this way will reproduce the
clustering properties of LBGs. In the right  panel of Fig.~\ref{fig9}
we compare the two-point correlation function of our ``starburst''
and ``massive-halo'' models. The correlation function   in the
starburst  model was computed from 120  different Monte-Carlo
realizations of catalogs of starbursting haloes. The correlation
lengths for the ``starburst'' and ``massive-halo'' models are
$2.5\hmpc$,  and $3.5\hmpc$, respectively.   The observed correlation
length is $r_0\bsim 3-4 \hmpc$ (Adelberger \etal 1998, Adelberger
2000, Arnouts \etal 1999, Giavalisco \&  Dickinson  2001).  The
massive halo picture is thus in somewhat  better agreement.  However,
the difference is small and as discussed above, Kolatt \etal (1999)
and Wechsler \etal (2001) find that they  can reproduce the observed
clustering properties if they select  ``colliding haloes'' as host
galaxies of starbursting  LBGs.  We thus  do not investigate this
issue any further here.

\subsection{Cosmic Variance}

\begin{figure} 
\mbox{\psfig{figure=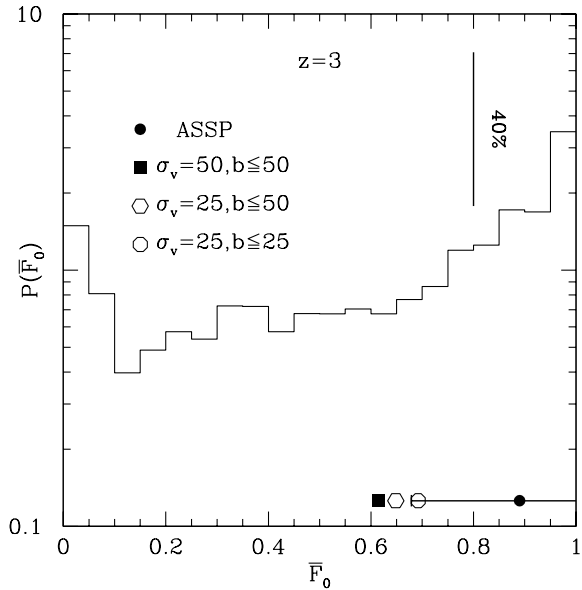,height=3.2in,width=3.2in}}
\caption{PDF of the  transmitted flux at separation $s\leq 0.5\hmpc$,
${\bar F}_0$, in the ``starburst'' picture. The filled circle is ${\bar
F}_0$ for $s\leq 0.5\hmpc$ as quoted by Adelberger \etal, whereas the
filled square is the ``starburst'' model.  Empty symbols show ${\bar
F}_0$ when the velocity shear and/or the  Doppler parameter around
haloes are reduced by a factor of two.}
\label{fig10}
\end{figure}

To understand better why the cosmic variance  error becomes so large
at small separation we plot the probability  distribution  $ P(\bar
F_0)$ in  Fig.~\ref{fig10}. The distribution is very broad.   
About 40\% of the haloes have ${\bar F}_0$ larger than 0.8.  
The probability of finding three
(uncorrelated) haloes with $\bar F\geq 0.8$ at separation $s\leq
0.5\hmpc$ is thus still  $\sim$5-10\%. The symbols at the bottom of 
Figure 5 show the increase of ${\bar F}_0$ if the thermal motions and
the  velocity shear are artificially reduced by a factor of two.  The
broad probability distribution also explains why the suggestion  of
Croft et. (2002) that LBGs are preferentially located in haloes with
a low-density  environment has such a strong effect on the  mean
transmission. 

\section{discussion}
\label{discussion}

We used mock spectra obtained from numerical simulations to
investigate the effect of fully ionised bubbles around the galactic
haloes hosting LBGS on the \hi\ distribution in the IGM as probed by
the \op\ forest in QSO spectra. By matching the probability
distribution and the power spectrum of the flux of mock spectra
obtained from a N-body simulation of a $\Lambda$CDM model with those
of observed spectra we derive an upper limit of $10\%$ on the volume
filling factor of bubbles. We further considered the mean transmitted
flux at a given distance from galactic haloes.  To model the
observational data we computed the mean transmitted flux $\bar F$ from
an annular average of the two-dimensional conditional flux ${\bar
F}(\pi,\sigma)$, in the same way as done by Adelberger et al.
(2003). As previous investigations we find that the observed decrease
of the transmission at separation $1\leq s\leq 5\hmpc$ to a halo is
well reproduced by the simulated spectra. It is due to the increased
matter density around haloes. We further find that an increase of the
transmissivity at the smallest separations $s\leq 0.5\hmpc$ requires
bubble radii of $\rw=1-1.5 \hmpc$. This is two to three times larger
than the separation at which the increase of the transmissivity is
observed. As discussed by Kollmeier \etal 2003 this is because in
redshift space the velocity shear of the material infalling onto the
halo fills in the cavity which is caused by the fully ionised wind
bubble in real space. For a model where the haloes are surrounded with
bubbles of size $\rw=1-1.5 \hmpc$ there is no effect on either the
flux PDF or the flux power spectrum.  The flux PDF and the flux power
spectrum of our simulated spectra are fully consistent with those of
observed spectra. This is because the volume filling factor of the
ionised bubbles is still small. We have investigated a variety of
scalings of bubble size with halo mass. We also compared a model where
LBGs are long-lived and are hosted by the most massive haloes with a
model where LBGs are short-lived starburts and are hosted by more
numerous less massive haloes.  Unlike Kolatt \etal (1999) and Wechsler
\etal (2001), we do not identify starbursts as ``colliding haloes''
but choose a fraction of less massive haloes at random. Matching the
observed transmission at scales $1\leq s\leq 5\hmpc$ is not difficult
and there is certainly more than one way to do so. Introducing
redshift errors at the expected level improves the agreement with the
observations.  The increase in transmission at small separation in our
model spectra spectra is, however, generally more gradual than that of
the observed measurements. As other authors we found it challenging to
match the observations at the smallest separation.  Numerical
simulations including galactic winds and simple analytical estimates
as the one in Appendix~\ref{cooltimescale} agree that the required
bubble sizes are difficult to achieve (Croft et al. 2002, Bruscoli et
al. 2002). We have thus particularly looked into models which are
consistent with the observed measurements and are most ``economical''
in terms of bubble size. The velocity shear around haloes increases
with increasing mass. Haloes of lower mass require thus smaller bubble
radii to explain the observed increase of the transmission at $s\leq
0.5\hmpc$. This may favour a model where LBGs are starburst galaxies
hosted by more numerous less massive haloes. We have taken special
care in estimating the errors on the flux near galaxies in a similar
way as done in the observations.  For small separations we found that
the expected error due to cosmic variance is about 30\% larger than
estimated by ASSP. Our starburst model is consistent with all
measurements for a bubble radius of $\rw=1 \hmpc$ and a volume filling
factor of 2\%.  A bubble size of $\rw=1 \hmpc$ appears, however, still
difficult to achieve for galactic winds.  If for some reason the
velocity shear around haloes is smaller than suggested by our (DM
only) simulation this would further reduce the required bubble size
and filling factor.  The correlation length for our ``starburst''
model is $2.5\hmpc$, slightly lower than the observations. Identifying
the LBGs with satellites infalling on high mass haloes, as in Kolatt
\etal (1999), would further improve the agreement with the observed
clustering properties of LBGs. However, this would be at the expense
of matching the observed small-scale flux $\bar F$, since the velocity
shear increases with halo mass.  Therefore we advocate the model in
which starbursts occur in small mass haloes. In a $\Lambda$CDM
cosmology of normalisation $\sigma_8=0.9$, this model yields a
reasonable match to the observed clustering of LBGs and transmitted
flux $\bar F$.  The mean transmitted flux of our simulated spectra at
the smallest separation is 30 percent smaller than that of the ASSP
sample. We would thus expect the observed mean transmission to
decrease substantially for larger samples if the model is correct.  A
sample with 12 LBGs at separation $s\leq 0.5\hmpc$ should reduce the
error to $\Delta{\bar F}\sim 0.1$. This should help to discriminate
against alternative models like the suggestion of Croft \etal (2002)
that LBGs are located preferentially in low-density environments.

\section{Acknowledgment}

We acknowledge stimulating discussions with J. Primack and M. Pettini.
We thank Juna Kollmeier, our referee, for constructive comments on the
manuscript.  This Research was supported by the Israel Science
Foundation (grant No. 090929-1), the German Israeli Foundation for
Scientific Research and Development, and the EC RTN network ``The
Physics of the Intergalactic Medium''.  VD would like to acknowledge
the MPA (Garching), the Institute of Astronomy (Cambridge) and the
Technion Computing Center where part of this work was accomplished.

\appendix

\section{Evolution of Bubbles}
\label{cooltimecale}

To explain the lack of \hi\ close to the haloes, the dilute, highly
ionised bubble has to survive for a sufficiently long time.  In this
appendix we present general order of magnitude estimates of the
wind/shock energetics, relate that to the  observed speeds, and
estimate cooling time scales.  We heavily rely on the work of Sedov
(1959), Weaver \etal (1977), Ostriker \& McKee (1988) and Tegmark
\etal (1993).

\subsection{Wind energetics}
\label{windenergetic}

The main characteristics of the wind, such as the temperature of the
inner plasma, the size $\rw$ of the bubble swept up by the shock and
the wind velocity $\vw$, depend on the total energy $E_{\rm w}$
released in the outflow. Neglecting energy losses $E_{\rm w}$ can be
estimated as
\begin{eqnarray}
E_{\rm w} &=&\frac{1}{2}M_{\rm w}\vw^2 \\ &\sim& 4.3\times
10^{56}\erg\,\left(\frac{\rw}{0.1\hmpc}\right)^3 \left(\frac{\vw}
{100\kms}\right)^2 \Delta_g \;, \nonumber
\label{eq30a}
\end{eqnarray} 
where $M_w$ is the gas mass inside $\rw$.  To get the numerical
estimation  we have assumed that $M_{\rm w}=(4\pi/3)\rw^3\rho_b
\Delta_g$ where $\rho_b$ is the mean background density and $\Delta_g$
is the density contrast inside $\rw$.   The characteristic radius $\rw
$ and velocity  $\vw$ of the wind are poorly constrained, but recent
observations suggest that $\rw\bsim 0.1\hmpc$ and $\vw\bsim 100\kms$
(Heckman \etal 2000, Pettini \etal 2001, Adelberger \etal
2003). On scales $r\sim 0.1\hmpc$ the value of $\Delta_g$
is typically about $2-5$, which gives $E_{\rm w}\sim 10^{57}\erg$.
This value is consistent  with the amount of energy $E_{SN}$ released
by supernovae during a burst of duration $t_{burst}\sim 10^7\yr$ and
of star formation rate $\dot{M}_\star\sim 10\myr$.

To assess whether the gas will be expelled from the halo one can apply
a simple binding argument (e.g. Dekel \& Silk 1986).  On the one hand
the potential energy $U$ of the gas which lies within the  halo is
$U\propto MM_{\rm g}/r_{\rm M}$, where $r_{\rm M}$ is the
characteristic radius of the halo, $r_{\rm M}\propto M^{1/3}$, and
$M_{\rm g}$ is the mass of the gas which lie within a radius $r_{\rm
M}$.  The mass of the gas in the halo is $M_{\rm g}\propto f_\star f_b
M$, where  $f_\star$ is the star formation efficiency and $f_b$ the
average fraction  of baryons in the halo. The potential energy is then
$U\propto M^{5/3}$.  On the other hand, if one assumes that winds are
driven by supernovae, the  energy released in the wind is $E_{\rm
w}\propto \dot{M}_\star t_{\rm burst}$, where $\dot{M}_\star\propto
M_{\rm g}$ is the star formation rate.  Hence, according to this crude
estimate, the ratio of the kinetic to binding  energy is $E_{\rm
w}/U\sim M^{-2/3}$.  We therefore expect $\rw$ to be a function of $M$
, and outflows to  escape more easily from low mass haloes (e.g. from
dwarf galaxies).  On this latter point, the measurements of Adelberger
\etal (2003) seem to  indicate that, if outflows are the cause of lack
of \hi\ absorption in the observed transmitted flux $\bar F$, they can
escape from high mass galaxies as well, and affect the IGM properties
on comoving scale as large as  $0.5\hmpc$.

\begin{figure} 
\resizebox{0.45\textwidth}{!}{\includegraphics{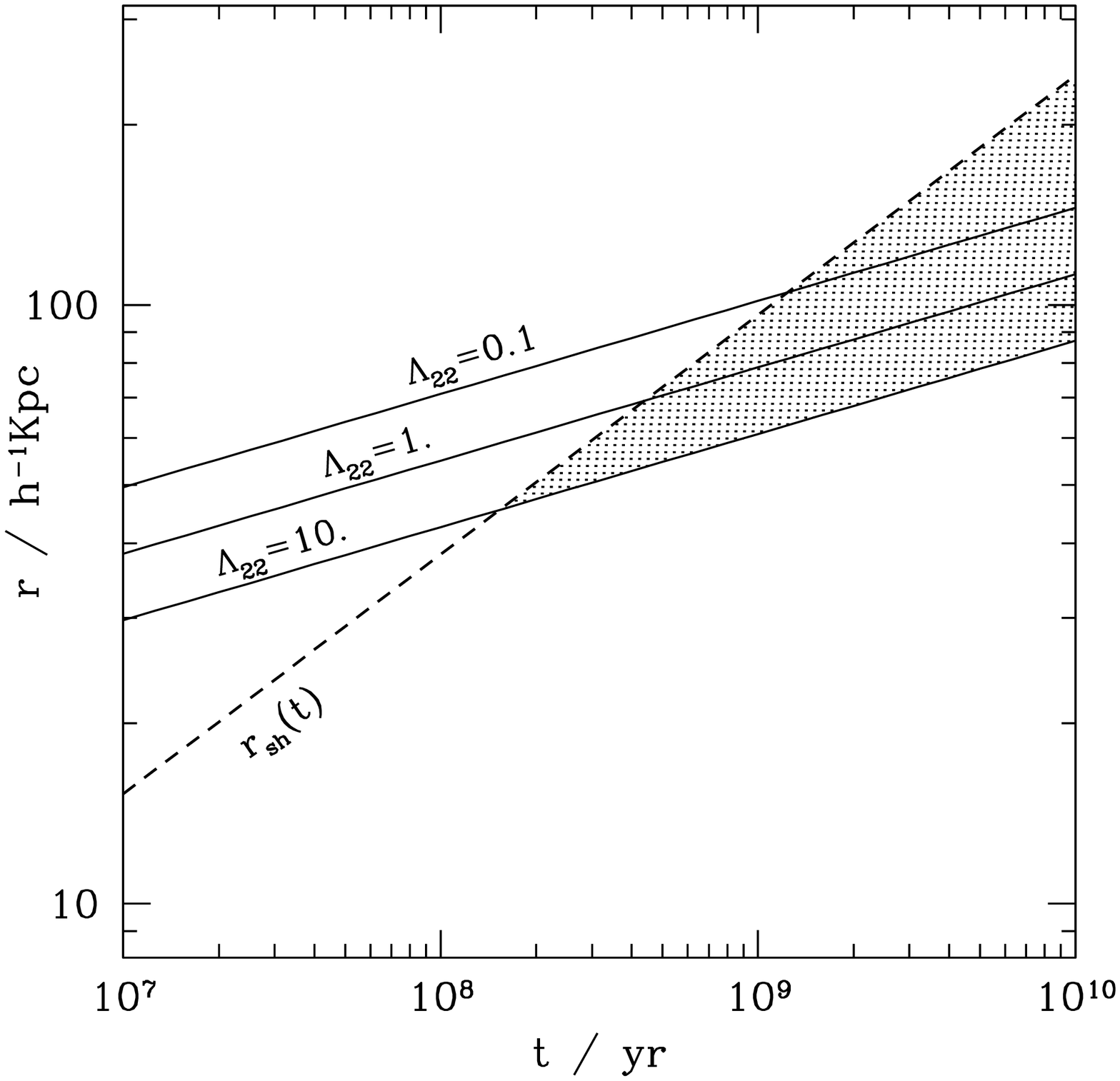}} 
\caption{ The cooling radius $r_{cool}(t)$ (solid curves) for various
cooling time rate.  The dashed curve is $r_{\rm sh}(t)$.  }
\label{figapp2}
\end{figure}

\subsection{Propagation of adiabatic shock}
\label{windpropagation}

A wind ejects energy in the IGM, driving a shock which heats up and
collisionally ionises the material it encounters.  We briefly  review
now the evolution of the shock front and the ionised bubble it
encompasses, before radiative losses become important.  We assume that
the gas follows the perfect gas law $P=(\gamma-1)u\rho$, where $P$,
$u$ and $\rho$ are, respectively, the pressure, the internal energy
and the density of the gas, and $\gamma$ is the ratio of specific
heats, which is $\gamma=5/3$ for a mono-atomic gas.  If the  expansion
of the universe can be ignored, the shock  is well described by a
strong adiabatic blast wave  which propagates into the IGM . In this
regime, accurate self-similar solutions exist.  Let $r_{\rm sh}(t)$ be
the position of the shock front at time $t$ after the burst. The jump
conditions relate the density $\rigm$ of the unperturbed IGM to the
density $\rho_{\rm sh}$, the pressure $P_{\rm sh}$ and the temperature
$T_{\rm sh}$ at the shock front. For a strong adiabatic shock the jump
conditions for pressure and density can be expressed as
\begin{equation}
\rho_{\rm sh}=\frac{(\gamma+1)}{(\gamma-1)}\rigm, ~~~
P_{\rm sh}=\frac{2}{\gamma-1}\rigm\frac{dr_{\rm sh}}{dt}^2.
\label{eq100}
\end{equation}
Hence, $\rho_{\rm sh}=4\rigm$ for an adiabatic index $\gamma=5/3$.
Assuming that the flow is self-similar, $r_{\rm sh}$ can be expressed
as a  function of the similarity variable $\xi$. The position of the
shock front corresponds then to a fixed value of $\xi\simeq 1$. Hence,
$r_{\rm sh}$ is  given  by (Sedov 1959)
\begin{equation}
r_{\rm sh}=\left(\frac{E_{\rm w}}{\rigm}\right)^{1/5}t^{2/5},
\label{eq101}
\end{equation}
where $E_{\rm w}\sim 10^{57}\erg$ is the energy released in the wind.
Inserting the value of $E_{\rm w}$ in the expressions of $r_{\rm sh}$ 
and $T_{\rm sh}$, we find
\begin{eqnarray}
r_{\rm sh}(t) &\simeq& 19\hkpc\left(\frac{t}{10^7\yr}\right)^{0.4} \\
& & \frac{1}{\Delta_g^{0.2}}\left(\frac{\om{b}h^2}{0.019}\right)^{-0.2}
    \left(\frac{E_{\rm w}}{10^{57}\erg}\right)^{0.2}
    \left(\frac{1+z}{4}\right)^{-0.6} \nonumber \\
T_{\rm sh}(t) &\simeq& 2.7\times 10^7\kel
    \left(\frac{t}{10^7\yr}\right)^{-1.2}\\
& & \frac{1}{\Delta_g^{0.4}}\left(\frac{\om{b}h^2}{0.019}\right)^{-0.4}
    \left(\frac{E_{\rm w}}{10^{57}\erg}\right)^{0.4}
    \left(\frac{1+z}{4}\right)^{-1.2} \nonumber
\label{eq102}
\end{eqnarray}
Since the temperature behind the shock front $T\propto P/\rho$ is a
decreasing function of $r$, $T$ reaches its minimum at the shock
front.  This solution holds only in the regime where the expansion is
negligible, i.e. $t\ll t_H$ where $t_H$ is the Hubble time
scale. Self-similar  solutions for $t>t_H$ including this latter
complication exist (Ostriker \& McKee 1988), but since energy and
momentum conservation break down for $t\sim t_H$, it is difficult to
express the asymptotic solution as a function of the initial $E_{\rm
w}$.

\subsection{Cooling time-scales and survival of bubbles}
\label{cooltimescale}

The evolution of such an expanding shock depends critically on cooling
which  will lower the temperature of the shell and its interior,
thereby increasing  the amount of \hi\ hydrogen available for \op
resonant scattering.  The expanding shell cools due to radiative
losses. The shock will  undergo a phase of shell formation which will
occur when the energy  dissipated in a volume $\propto r_{\rm
sh}(t)^3$ will be of the order of $E_{\rm w}$.

Cooling is expected to be more efficient at the shock front since, at
$r=r_{\rm sh}$, the density (temperature) is larger (lower) than  at
$r<r_{\rm sh}$. Thus, shell formation first occurs at the shock front.
As long as we are in the regime of low density, $\rho\ll \rigm$, and
high  temperature, $T\gg 10^6\kel$, radiative cooling is dominated by
thermal  bremsstrahlung.  However, for temperature $T\lsim 10^6\kel$,
atomic line cooling might  contribute significantly to the cooling
rate $\Lambda$ (in $\ergscms$)  if the wind is enriched with heavy
elements.  For a wind of metallicity $Z\sim 0.1 Z_\odot$
(e.g. Adelberger \etal 2003),  the cooling rate is $\Lambda\sim
10^{-22}\erg$ in the range  $10^4\lsim T\lsim 10^6\kel$
(e.g. Sutherland \& Dopita 1993).

We assume that cooling at the shock front, where the temperature is
$T\lsim 10^6\kel$, is dominated by atomic line transitions. This is
true when the temperature of the shell is $T_{\rm sh}\lsim 10^6\kel$,
i.e. for times $t\bsim 10^8\yr$. The cooling time scale at the shock
front is then $t_{cool}(r_{\rm sh},t)=3k_B T_{\rm sh}/(2n_{\rm
sh}\Lambda)$,
\begin{eqnarray}
t_{cool}(r_{\rm sh},t)&=&1.9\times 10^{10}\yr 
    \left(\frac{t}{10^7\yr}\right)^{-1.2}\Lambda_{22}^{-1} \\ 
& & \frac{1}{\Delta_g^{1.4}}\left(\frac{\om{b}h^2}{0.019}\right)^{-1.4}
    \left(\frac{E_{\rm w}}{10^{57}\erg}\right)^{0.4} \left(\frac{1+z}{4}
    \right)^{-4.2}\nonumber \;,
\label{eq103}
\end{eqnarray}
where the cooling rate $\Lambda_{22}$ is in unit of $10^{-22}\ergscms$ 
and is assumed to be constant in the temperature range 
$10^4\lsim T_{\rm sh}\lsim 10^6\kel$.  

Once the shell starts forming, the temperature $T_{\rm sh}$ of the
shell cools  to $\sim 10^4\kel$ on a very short time scale.  However,
in the range $T\lsim 10^4\kel$, the cooling rate $\Lambda$  drops
sharply from $\sim 10^{-22}$ down to $\sim 10^{-26}\ergscms$
(Sutherland \& Dopita 1993) and the shell cannot cool  below
temperature $T\sim 10^4\kel$. We define now a radius $r_{cool}(t)$
such that the cooling time scale $t_{cool}$ as a function of radius
$r$ satisfies $t_{cool}<t$ for $r>r_{cool}$. According to this
definition, the gas  between $r_{cool}(t)$ and $r_{\rm sh}(t)$ has
cooled  onto the shell by the time $t$.  To express $t_{cool}$ as a
function of $r$ and $t$, it should be noted that the number density
$n_{\rm w}(r,t)$ behind the shock front (in $\pcc$) scales as $n_{\rm
w}(r,t)= (r/r_{\rm sh})^\alpha n_{\rm sh}(t)$ with $\alpha\sim 4.5$
typically. This relation is however valid in the asymptotic regime
only.  Therefore, assuming that the plasma has reached pressure
equilibrium,  the temperature $T_{\rm w}(r,t)$ behind the shock front
is  $T_{\rm w}= (r_{\rm sh}/r)^\alpha T_{\rm sh}$, and the radiative
cooling time scale $t_{cool}$ is
\begin{equation}
t_{cool}(r,t)=\left(\frac{r_{\rm sh}}{r}\right)^{2\alpha} 
t_{cool}(r_{\rm sh},t)\;. 
\label{eq105}
\end{equation} 
On Fig.~\ref{figapp2} the solid curves are $r_{cool}(t)$ for various
cooling time rate $\Lambda_{22}$, and the dashed curve is $r_{\rm
sh}(t)$. The shell starts forming at the shock front when
$r_{cool}(t)$ intersects $r_{\rm sh}(t)$. For larger time, cooling
occurs for $r_{cool}(t)<r<r_{\rm sh}(t)$ (shaded area).  Since most of
the gas swept up by the wind lies at $r\lsim r_{\rm sh}$,  a large
fraction of gas cools down to $T\sim 10^4\kel$ shortly after the shell
formation. For $\Lambda_{22}=1$, about 30\% of the mass enclosed in
the bubble has cooled into the shell over a Hubble time $t_H$, where
$t_H\simeq 1.2\times 10^9\left(\om{m}h^2\right)^{-1/2}$ at $z=3$.
For a high metallicity wind $Z\bsim 0.1 Z_\odot$, cooling becomes
severe for $t\bsim$ a few $10^8\yr$, a time scale shorter than the
Hubble time scale $t_H$ at $z=3$. On the contrary, for a low
metallicity wind, $Z\ll 0.1 Z_\odot$, the cooling rate is
$\Lambda_{22}\ll 1$ and cooling will not become significant for
$t\lsim t_H$.  Hence, most of the bubbles will not survive over a
Hubble time scale unless the cooling time rate $\Lambda_{22}$, or the
wind metallicity $Z$, is very low.  For radii $r\lsim r_{cool}(t)$,
the temperature is higher than $10^6\kel$, and thermal bremsstrahlung
and inverse Compton cooling prevail over line cooling, but the
corresponding cooling times are still smaller than the Hubble time 
at $z=3$.

\subsection{Absorption by the cold shell}

To compute the thickness $\Delta s(t)$ of the shell, we assume that
the pressure is constant through the inner plasma and the shell. Since
the temperature of the shell, $T_{\rm sh}\sim 10^4\kel$, is much lower
than the overall temperature $T_{\rm w}\bsim 10^6\kel$ of the highly
ionised  plasma, we expect $\Delta s(t)\ll r_{\rm sh}(t)$, i.e. the
shell is thin.  Taking into account this assumption, we find
\begin{equation}
\frac{\Delta s}{r_{\rm sh}}=\frac{1}{3}\left(\frac{T_{\rm sh}}{T_{\rm w}}
\right)\left[1-\left(\frac{r_{cool}}{r_{\rm sh}}\right)^3\right]\;,
\label{eq106}
\end{equation}
a result consistent with a thin shell approximation. Assuming 
$T_{\rm w}=100T_{\rm sh}=10^6\kel$ and neglecting the expansion
 of the Universe gives $\Delta s\lsim 0.01r_{\rm sh}$. 
$\Delta s$ should thus not  exceed a few $\hkpc$, a scale which 
cannot be resolved in our simulation.  
The optical depth for resonant Ly$\alpha$
scattering is given by 
\begin{eqnarray}
\tau &\sim& \sigma_0 n_{\rm sh}\Delta s \\ &\sim& 1.6\,
\eta\Delta_g\left(\frac{\om{b}h^2}{0.019}\right) \left(\frac{1+z}{4}
\right)^2\left(\frac{\rw}{0.1\hmpc}\right)\nonumber \;,
\label{eq107}
\end{eqnarray}
where $\eta\simeq 1$ is the \hi\ neutral fraction at temperature
$T\sim 10^4\kel$.  \op absorption is expected  for reasonable values
of the density contrast $\Delta g\sim 2-3$, the typical comoving
radius of the bubble  $\rw\sim 0.1\hmpc$, and the baryon fraction
$\om{b}h^2\sim 0.02$.  However, since we cannot resolve the thin shell
in our simulation, we have nevertheless ignored its \op absorption.

\section{Numerical convergence}
\label{convergence}

\begin{figure} 
\resizebox{0.45\textwidth}{!}{\includegraphics{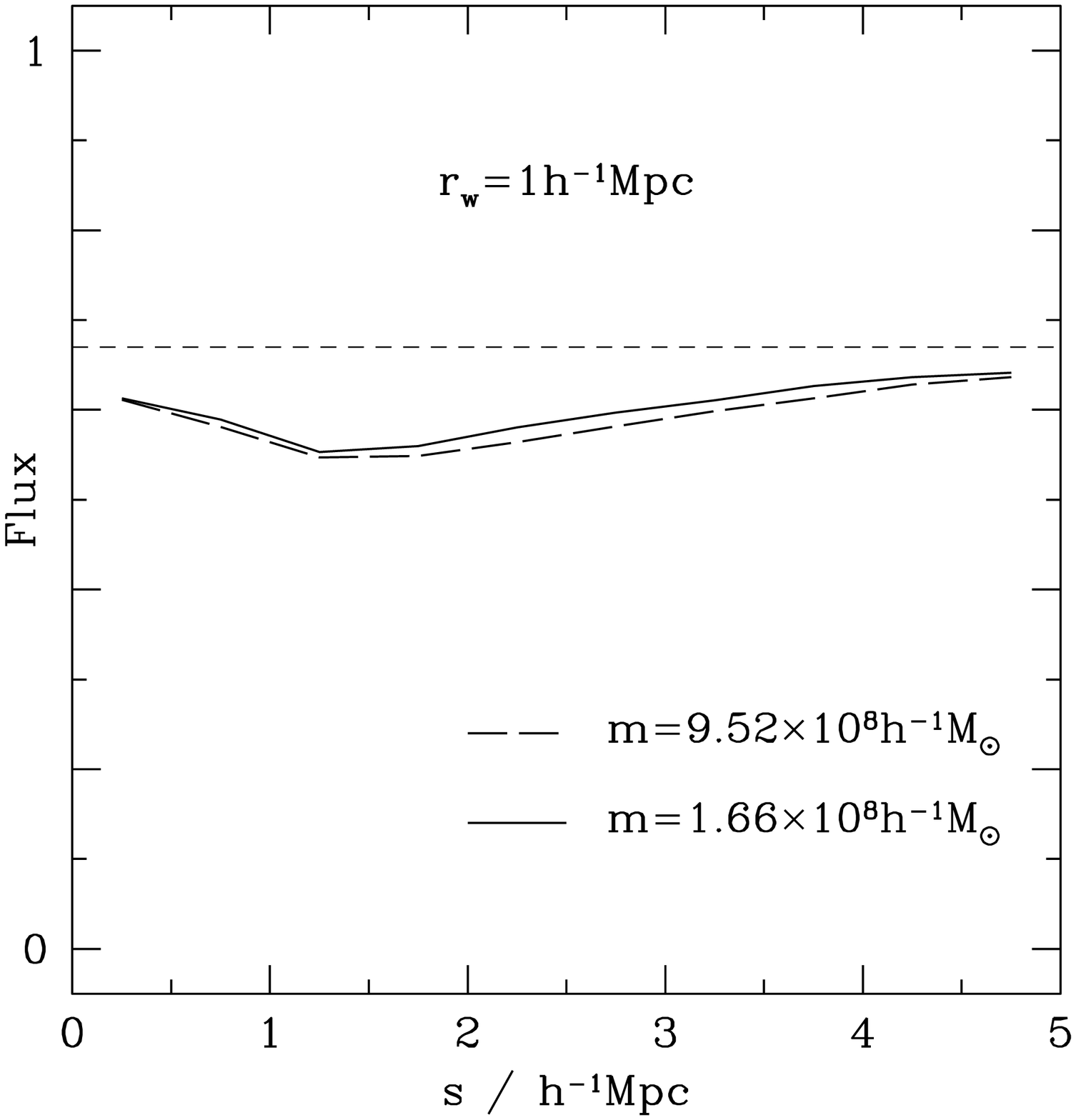}} 
\caption{ The transmitted flux $\bar F$ in the ``starburst'' scenario
with bubbles of radius $\rw=1\hmpc$, with DM particle mass resolution
$m=9.52\times 10^8$ (dashed curve) and $1.66\times 10^8\mdh$ (solid
curve). The horizontal dashed line is the flux normalization, $\la
F\ra=0.67$.  }
\label{figapp3}
\end{figure}

To assess the sensitivity of the transmitted flux to numerical
resolution, we perform a calculation with the highest resolution
simulation of the series, which has a mass resolution $m=1.66\times
10^8\mdh$. The halo catalog and the flux in the simulation are
computed as in Section~\S\ref{nbody} with a pixel resolution
$\Delta=60\hkpc$ (comoving).   We added a Gaussian deviate of variance
$\sigma_v=150\kms$ to the halo positions.  In Fig.~\ref{figapp3} we
show the transmitted flux as computed in the ``starburst'' scenario
for a bubble radius $\rw=1\hmpc$. The mass resolution is $m=9.52$
(dashed curve) and $m=1.66\times 10^8\mdh$ (solid curve).  In the
latter the strong absorption lines are better resolved, thereby
slightly reducing the normalization of the optical depth ${\cal A}$.
Therefore, the optical depth of the absorbing material is lower, and
the transmitted flux is slightly larger, by about $\Delta{\bar F}\sim
0.02$ on scales of $2\leq s\leq 4\hmpc$.

\end{document}